\documentclass[sigconf]{acmart}
\acmConference[ICSE 2024]{46th International Conference on Software Engineering}{April 2024}{Lisbon, Portugal}

\usepackage{amsmath}
\usepackage{xcolor}
\usepackage{color}
\usepackage{listings}
\usepackage{fixltx2e}
\usepackage{tabularx}
\usepackage{balance}
\usepackage{float}
\usepackage{graphicx}
\usepackage{algorithm}
\usepackage[noend]{algpseudocode}
\usepackage{hyperref}
\usepackage{subfig}

\definecolor{coralred}{rgb}{1.0, 0.25, 0.25}
\definecolor{bostonuniversityred}{rgb}{0.8, 0.0, 0.0}
\definecolor{cadmiumred}{rgb}{0.89, 0.0, 0.13}
\definecolor{darkcandyapplered}{rgb}{0.64, 0.0, 0.0}
\definecolor{britishracinggreen}{rgb}{0.0, 0.26, 0.15}
\definecolor{coolblack}{rgb}{0.0, 0.18, 0.39}
\definecolor{bittersweet}{rgb}{1.0, 0.44, 0.37}
\definecolor{bubblegum}{rgb}{0.99, 0.76, 0.8}
\definecolor{persianred}{rgb}{0.8, 0.2, 0.2}
\definecolor{amber}{rgb}{1.0, 0.49, 0.0}

\makeatletter
\renewcommand{\ALG@beginalgorithmic}{\footnotesize}
\makeatother

\algrenewcommand\algorithmicindent{0.5em}
\algrenewcommand\alglinenumber[1]{\tiny#1}

\algnewcommand\algorithmicof{\textbf{of}}
\algnewcommand\algorithmiccase{\textbf{case}}
\algdef{SE}[CASE]{Case}{EndCase}[1]{\algorithmiccase\ #1 \algorithmicof}{\algorithmicend\ \algorithmiccase}%
\algtext*{EndCase}%

\algnewcommand\algorithmicinstrument{\textbf{instrument}}
\algdef{SE}[INSTRUMENT]{Instrument}{EndInstrument}[2]{\algorithmicinstrument\ \textsc{#1}\ \textsc{\textbf{#2}}}{\algorithmicend\ \algorithmicinstrument}%
\algtext*{EndInstrument}%

\algnewcommand\algorithmicrelation{\textbf{relation}}
\algdef{SE}[RELATION]{Relation}{EndRelation}[1]{\algorithmicrelation\ #1}{\algorithmicend\ \algorithmicrelation}%
\algtext*{EndRelation}%

\algnewcommand\algorithmicabsrelation{\textbf{abstract relation}}
\algdef{SE}[ABSRELATION]{AbsRelation}{EndAbsRelation}[1]{\algorithmicabsrelation\ #1}{\algorithmicend\ \algorithmicabsrelation}%
\algtext*{EndAbsRelation}%

\algnewcommand\algorithmicabsfunction{\textbf{abstract function}}
\algdef{SE}[ABSFUNCTION]{AbsFunction}{EndAbsFunction}[1]{\algorithmicabsfunction\ #1}{\algorithmicend\ \algorithmicabsfunction}%
\algtext*{EndAbsFunction}%

\def\fcall#1{\textsc{#1}}

\def\var#1{\texttt{#1}}

\def\lb{\langle}
\def\rb{\rangle}

\hypersetup{
    colorlinks=true,
    linkcolor=blue,
    urlcolor=cyan,
}

\def\tensorflow{\textsc{TensorFlow}}
\def\keras{\textsc{Keras}}
\def\bert{\textsc{BERT}}
\def\albert{\textsc{ALBERT}}

\def\oursys{\textsc{Smaragdine}}

\def\BibTeX{{\rm B\kern-.05em{\sc i\kern-.025em b}\kern-.08em
    T\kern-.1667em\lower.7ex\hbox{E}\kern-.125emX}}

 \begin{abstract}

With the rapid growth of Artificial Intelligence (AI) applications supported by deep learning (DL), the energy efficiency of these applications has an increasingly large impact on sustainability. We introduce \oursys{}, a new energy accounting system for tensor-based DL programs implemented with \tensorflow{}. At the heart of \oursys{} is a novel \emph{white-box} methodology of energy accounting: \oursys{} is aware of the internal structure of the DL program, which we call \emph{tensor-aware energy accounting}. With \oursys{}, the energy consumption of a DL program can be broken down into units aligned with its logical hierarchical decomposition structure. We apply \oursys{} for understanding the energy behavior of \bert{}, one of the most widely used language models. Layer-by-layer and tensor-by-tensor, \oursys{} is capable of identifying the highest energy/power-consuming components of \bert{}. Furthermore, we conduct two case studies on how \oursys{} supports downstream toolchain building, one on the comparative energy impact of hyperparameter tuning of \bert{}, the other on the energy behavior evolution when \bert{} evolves to its next generation, \albert{}.

\end{abstract}

\begin{document}

%\title{Identifying Energy Hotspots in \tensorflow{} Applications}

\title{Tensor-Aware Energy Accounting} % in \tensorflow{} Applications}

\author{Timur Babakol and Yu David Liu}
\affiliation{
    SUNY Binghamton
    \city{Binghamton}
    \state{NY}
    \country{USA}
}
\email{{tbabako1,davidl}@binghamton.edu}

\maketitle

\section{Introduction}
\label{sec:intro}

% \dnote{I have some second thoughts on whether the "nested DNNs" view is correct. Looking at the path locations of the tensors, it feels more like composing tensors through programs. To some extent, the tensor-composed program view is better for an SE conference, because if we say what we are doing is just a DNN, then reviewers think we should just submit to ICML or ACL. Just like PL people, SE people like *programs*. I have changed nearly all occurrences of DNNs to "DL programs". I will also update Sec 2 to reflect this view. Please reread. }

%With Machine Learning (ML) applications being an integral part of computing, 
Green AI~\cite{deep-learning-e-policy} is a fundamental challenge with far-reaching implications on the future practice of software engineering, and the sustainability of our society~\cite{NSFvmware}. Deep learning (DL) ~\cite{Linnainmaa1976TaylorEO, fukushima1980neocognitron, bengio2000neural, collobert2008unified, rumelhart1985learning} --- the technology that drives the current wave of AI revolution --- happens to be excessively energy-hungry. 
%they demand power-hungry heterogeneous computing platforms~\cite{language-models-are-few-shot-learners}, and training alone can take days if not months~\cite{Alzubaidi2021ReviewOD}. 
%For example, the training of 
For example, training DL-based large language models is estimated to consume 1,287 megawatt hours of electricity~\cite{patterson2021}. 
%which is more than 26500 years of electricity use by an average household. IT sectors continue to project an increase of electricity us for ML workloads~\cite{sustainableai}.
%Power-hungry ML applications are a significant workload of modern data centers, whose energy impact is of grave concern~\cite{deep-learning-e-policy}. Operationally, excessive heat dissipation due to extended high-power program executions may destabilize hardware reliability, which requires expensive thermal management and cooling to offset.  
% To either optimize energy consumption or fix potential energy bugs, 
Optimizing the energy consumption of DL systems and applications is a fast-developing direction with intense interest~\cite{delight2016, neuralpower, GARCIAMARTIN201975}.

The first step toward change is awareness~\footnote{Nathaniel Branden, The Psychology of Self-Esteem, 1969}. Underpinning many solutions of energy optimization is the fundamental problem of \emph{energy accounting}: a deep understanding of energy consumption by breaking it down in (software or hardware) components. This is in contrast with the ``monolithic'' \emph{black-box} approach, e.g., measuring the end-to-end energy consumption of a DL training session. Energy accounting is a classic problem, with established solutions~\cite{icount, currentcy, ecosystem, chappie} focusing on breaking down the energy consumption by hardware components and OS system components.  It is not difficult to envision a \emph{gray-box} approach that retrofits these existing solutions to DL programs, i.e., breaking down the energy consumption of a DL training session by architectural units or OS threads.

%consumption of ML applications. Indeed, systems-level energy profiling tools have long existed~\cite{RAPL, jrapl, Babakol:2020, eflect}, which can effectively break down the energy consumption of an application per hardware component (\emph{e.g.}, CPU, GPU, memory), or per OS scheduling unit (\emph{e.g.}, process, thread). While such tools can provide useful insights on the energy behavior of ML applications, they are \emph{blind} to the \emph{machine learning} nature of these applications. 

% key insight is we can perform white box analysis

Our key insight is that, be it a black-box approach or a gray-box approach, it is a missed opportunity that an energy accounting system ignores the \emph{unique abstractions latent in the DL program}. After all, a DL program is highly structured, often broken down in modules formed by layers of tensors. Both the black-box approach and the gray-box approach pessimistically treat the DL program \emph{just like any other program}. \emph{How} can we leverage the structural information in the DL program for energy accounting? \emph{What benefits} do we gain with this new flavor of energy accounting? % This paper is a quest for answers to these questions. 

%In our view, The ML-blind approach to energy accounting is a missed opportunity, and it is practically undesirable in practice. For ML applications based on tensors, higher-level programming abstractions --- beyond systems-level abstractions such as GPU kernels or OS threads --- have long existed. Indeed, the tensor consists of \emph{neurons} connect with data flows. An energy profile that says the GPU accounts for 80\% of energy consumption, even with a per-thread breakdown, reveals little on the itensorerworkings of the tensor. The impedance mismatch between the abstraction understood by a programmer and that reported by a profiler has consequence in optimization and debugging. The programmer may be fully aware a particular GPU thread may be an ``energy hog'', but how does she map that knowledge to neuron-based computation in her tensor? 

\subsection{Tensor-Aware Energy Accounting}

%\dnote{or should it be neuron-aware? "network" can mean many things it's true.}

%\tnote{layer or operation is the most appropriate because the energy is mapped to a tensor operation.}

In this paper, we introduce \oursys{}~\footnote{Smaragdine means "pertaining to emeralds" in Latin and refers to the Smaragdine Tabula or Emerald Tablet, a legendary alchemical text that described the creation of the Philosopher's Stone.}, a novel \emph{multi-grained} energy accounting system for \tensorflow{}~\cite{tensorflow}-based DL programs that \emph{aligns the decomposition of energy consumption with that of the logical structure of the DL program}. At the heart of \oursys{} is a novel \emph{white-box} methodology of energy accounting: the system is aware of the internal structure of the DL program, and its energy consumption can be broken down following its abstraction boundaries. The output of \oursys{} is a series of nested \emph{Energy Distribution Diagrams} (EDD) corresponding to the \emph{hierarchical decomposition} of the DL program. For instance, a top-level EDD shows how the overall energy consumption of a DL program is broken down to modules; the EDDs for sub-modules show how the energy consumption of a particular module is broken down to Deep Neural Network (DNN) layers; and so on. With the hierarchically decomposed EDDs, the atomic unit of energy accounting is a tensor operation in \tensorflow-based implementations. 

%Concretely, our new system,  is designed for ML applications built with \tensorflow{}~\cite{tensorflow}, a software framework widely used for tensor-based machine learning. With \tensorflow{}, an tensor is arranged as a stateful datatflow graph, organized into a hierarchy of \emph{layers}. \oursys{} is able to produce an energy (or power) profile at the operational granularity of the tensor, such as per layer, or per module, while retaining the flow. With our tool, a developer can identify energy hotspots in her model and understand the tensor's runtime behavior.

The white-box approach of \oursys{} has some distinct benefits. First, it complements the active research of explainable AI ~\cite{tensorflow, sagemaker, pace} with a perspective on the \emph{explainability} of non-functional properties such as energy consumption. The hierarchical decomposition structure of EDDs explains how energy consumption is distributed in a layer-by-layer, or even tensor-by-tensor manner. Second, EDDs --- with their ``logical'' nature of energy accounting --- may offer insights to downstream designs in energy optimization and energy debugging. With EDDs, it is a trivial task for a downstream designer to zero in on the ``energy hotspot'' --- the highest energy-consuming unit (a module, a layer, or even a tensor) of the DL program. These energy hotspots are likely to be the ideal candidates for optimization or bug fixes due to their proportionally larger impact. Third, the white-box approach of \oursys{} promotes the study of \emph{portability} in energy accounting. In gray-box approaches, the result of energy consumption is fundamentally platform-specific: the breakdown of energy consumption across hardware components is dependent on what hardware components each platform has. In contrast, \oursys{} breaks down the energy consumption through logical components of a DL program. 

%when the same \tensorflow{}-based ML application is deployed on two machines with different configurations, the physical energy consumption breakdowns exhibited by two deployments are likely to be radically different, but is there any energy characteristics inherent to the logical structure of the tensor itself that remains stable independent of deployment environments?

The design of \oursys{} must overcome two major challenges. First, there remains a \emph{semantic gap} between the DL program and the underlying physical system: no ready-made profiler or tool can connect the semantic features of the DL program execution to energy consumption. The solution of \oursys{} is a \emph{trace-based alignment} algorithm, which conceptually can be viewed as a monitor that simultaneously tracks the DL program runtime events and the energy consumption of the system, and aligns the two to compute the EDD (details in \S~\ref{sec:design}). Second, there are complex interactions at the {\it application-system interface}. For example, most \tensorflow{} applications are multi-threaded running on heterogeneous platforms, i.e. with both CPUs and GPUs. How to account for energy consumption is non-trivial (see \S~\ref{sec:design} for details).

\subsection{Understanding \bert{} Energy Behavior}

To evaluate the effectiveness of tensor-based energy accounting, we apply \oursys{} to \bert{}~\cite{devlin-etal-2019-bert}, a widely used text analysis model. In the domain of natural language processing (NLP), \bert{} plays a central role in powering numerous NLP end-user applications. In \S~\ref{sec:evaluation}, we show how the \bert{} application can be hierarchically decomposed into EDDs. We also show how \bert{} transformers~\cite{transformers} --- especially its attention modules --- dominate the energy consumption in different stages of \bert{} use, from pre-training to fine-tuning to prediction. Throughout experiments, we find \oursys{} incurs low overhead while retaining high precision and stability.

To further demonstrate the usefulness of \oursys{} in building the downstream toolchain, we use \bert{} in two case studies. First, we provide a white-box study on the impact of hyperparameter tuning of \bert{}. We show the energy/power consumption trends with different configurations in the number of layers and the number of hidden embeddings. Second, we conduct an evolutionary study to compare \bert{} with a newer variant, \albert{}~\cite{Lan2020ALBERT:}. Through \oursys{}, we show how the energy behavior has evolved from \bert{} to \albert{}.

%Indeed, many variants of \bert{} algorithms exist \cite{Lan2020ALBERT:, liu2020roberta}. In this sense, our \bert{}-based case study is over a large \emph{family} of tensors with a common NLP goal. Thanks to \oursys{}, we are capable of understanding the energy behavior of different \bert{} variants w.r.t. not only individual operation but also over large hierarchies. 

%Our design is not unique only to \bert{} and works with event traces produced from other \tensorflow{} executions through the \texttt{timeline} library.

\subsection{Contributions}

To the best of our knowledge, \oursys{} is the first white-box energy accounting system for tensor-based DL programs. 
%\oursys{} presents designers a bridge between their expectations and the bare metal the network runs on. 
This paper makes the following contributions:

\begin{itemize}
    \item the distinct methodology for the energy accounting of DL programs where energy accounting is aware of the internal structure of the DL program, i.e., \emph{semantics-aware}, illustrated by EDDs (see \S~\ref{sec:design}) %and Tensor Energy Footprint (TEF, see \S~\ref{sec:design}). 
    \item a trace-based alignment algorithm for bridging the semantic gap of energy accounting while considering complex application-system interactions (see \S~\ref{sec:design})
    
    %accounting energy of event traces
    %\item multi-device aware application to system mapping
    %\item a designer-guided recomposition of accounting that can span over many sub-network hierarchies

    \item an in-depth evaluation of \oursys{} through \bert{}, revealing the module-level, layer-level,  and tensor-level energy behavior of \bert{} (see \S~\ref{sec:evaluation})
    \item a comparative study on the energy/power impact of hyperparameter tuning in \bert{}, and the energy/power behavior evolution from \bert{} to its next generation, \albert{} (see \S~\ref{sec:variants})
\end{itemize}

\oursys{} is an open-source project. The source code and all raw data of this paper can be found at the anonymous site: {\tt \url{https://github.com/project-smaragdine/smaragdine}}.

%While this is just a case study, we believe \oursys{} is a unique contribution to sustainability in deep learning. The presented methodology is aware of the entire network topology, as opposed to specific families of models. Our recomposition seeks to transform the network into an easily consumable footprint despite the enormous size of a model like \bert{}.

\section{Background}
\label{sec:background}

% We now provide a primer on neural networks, especially deep learning networks. In addition, we will present the challenges associated with energy accounting as a result of the abstractions provided by deep learning.

\paragraph{Deep Learning Programs}
% \dnote{turn them into backslash paragraph}
A DL program is a dataflow program that composes a number of \textit{neural networks} (NNs) --- and often \textit{deep neural networks} (DNNs) --- together. 
A \textit{neural network} (NN) is a collection of \emph{neurons} wired together, where each neuron serves as a transformation function. Each neuron can be associated with a \emph{weight}, a potentially adjustable value that indicates the importance of the transformation. 
%Neural networks (NN) are a class of learning models that wire together smaller models. Classically, the primitive unit of an NN is a neuron, a (potentially \textit{adjustable}) transformation function. A collection of neurons can be wired together into a \textit{neural network}. 
DNNs hierarchically organize NNs together, each of which is called a \emph{layer}. % -- an abstraction of the network's connections. 
Each DNN layer can be implemented by a tensor~\cite{novikov2015tensorizing}, which generalizes scalar and vector computations over the input/output of the neurons.

Semantically, one may view a DL program as a potentially nested dataflow program. A DL program consists of \emph{modules} wired together through dataflows, and each module can be implemented either as a tensor, or another (nested) dataflow program. A concretization of this view is to consider a DL program as a ``nested DNN,'' consisting of layers wired together. Each layer may either be an atomic \emph{tensor layer} or \emph{composite layer}, i.e. another DNN. In the rest of the paper, we adopt this view.

% To avoid confusion, we henceforth name a layer implemented by another DNN as an \emph{composite layer}, and by a tensor as a \emph{tensor layer}. \tnote{the end sounds redundant}

%In realistic systems, the computation that happens within a layer can be complex, and hence often implemented by another DNN. With that, DNNs are \emph{nested}: a DNN may consist of another DNN as the implementation of some of its layers. 

DL programs can be \textit{trained}, i.e., adjusting the neuron weights of its resident NNs to fit a data set. Once trained, a DL program can be used for \textit{prediction}, i.e., estimating an output from an input. DNNs are trained through \textit{backpropagation} \cite{Linnainmaa1976TaylorEO, rumelhart1985learning}, where the error of the model's prediction is used to adjust weights. This requires computation of the network's \textit{gradient} -- the differential impact of neuron connections -- to correctly adjust the weights.

\paragraph{\bert{}}
\bert{}~\cite{devlin-etal-2019-bert} is a tunable NLP model relying on DNNs. \bert{} popularized the idea of splitting training into two stages: \emph{pre-training} and \emph{fine-tuning}. 
\bert{} is initially \textit{pre-trained} on a general data set -- where all model weights are trained for a long time. \bert{} can then be \textit{fine-tuned} on a curated data set -- where only a subset of model weights are trained for a shorter time. As a result, \bert{} can be quickly configured to solve specific kinds of text problems without retraining from scratch~\cite{devlin-etal-2019-bert, patent-bert, bio-bert}, leading to a break-through in NLP.

%One of the critical breakthroughs for DNNs is \bert{}~\cite{devlin-etal-2019-bert}, a tunable NLP model. 

For NLP models, two important tasks are \emph{embedding} and \emph{encoding}. Embedding represents the input (such as a text) for processing, and encoding addresses the transformation between the input and the output.  In design, \bert{}'s encoding module, called an \emph{encoder}, is a nested DNN: it stacks together a number of modules (i.e., composite layers) each of which is called a \textit{transformer}~\cite{transformers}. For our purpose, note that a transformer is a nested dataflow program with NNs inside. A key \bert{} innovation is one of the NNs nested inside a transformer, called  \textit{self-attention}. This unit enables a (now widely used) form of encoding known as bidirectional representation.

\paragraph{\tensorflow{}}

\tensorflow{} \cite{tensorflow} is a machine learning library that supports complex tensor calculus. \tensorflow{} programs can be designed using the \keras{} API, a framework for constructing DL programs. For DL training, computing the gradient by hand is non-trivial. To overcome this, \tensorflow{} automates this process by using \textit{automatic differentiation} \cite{Linnainmaa1976TaylorEO}.

\section{\oursys{} Design}
\label{sec:design}

% \oursys{} combines a \tensorflow{} runtime and application-level energy by aligning device measurements to regions of time. An algorithm specification is shown in Algorithm~\ref{alg:smaragdine_accounting}. For readers unfamiliar with the execution of tensor-based deep learning systems, we first provide a background.

\subsection{Problem Statement}

We use $ l \in \mathbb{LN}$ represent layer names, $t \in \mathbb{TN}$ tensor (layer) names, and $c \in \mathbb{CN}$ composite layer names. $\mathbb{LN} = \mathbb{TN} \cup \mathbb{CN}$. We further use $\mathbb{TO}$ to represent the set of tensor operations. 

\begin{definition}[DL Program]
\label{def:dlp}
We define a program $P \in \mathbb{P}$ as a directed graph $\lb N; E \rb$ where $\emph{N} =  \emph{C} \cup \emph{T}$, and $\emph{C}: \mathbb{CN} \rightharpoonup \mathbb{P}$ is a bijective partial function for the set of composite layers, $\emph{T}: \mathbb{TN} \rightharpoonup \mathbb{TO}$ is a bijective partial function for the set of tensor layers, and $E: \mathbb{LN} \rightharpoonup \mathbb{LN}$ is a partial mapping denoting the dataflow among layers. % \dnote{hmm, in DNNs, can one output be fed into multiple inputs? Can one input be fed to multiple outputs? }
\end{definition}

%\tnote{is \mathbb{O} supposed to be \mathbb{TO}?}

The primary goal of \oursys{} --- i.e., tensor-level energy accounting --- is to produce an EDD: 

\begin{definition}[Energy Distribution Diagram]
\label{def:edd}
We define an $\emph{EDD} \in \mathbb{EDD}$ as a directed graph  $\lb \mathcal{N}; E \rb$ where $\mathcal{N} =  \mathcal{C} \cup \mathcal{T}$, and $\mathcal{C}: \mathbb{CN} \rightharpoonup \mathbb{EDD}$ is a partial function for the set of EDDs (indexed by composite layer names), $\mathcal{T}: \mathbb{TN} \rightharpoonup \mathbb{ENERGY}$ is a partial function for the set of energy consumption values (indexed by tensor names). 

\end{definition}

It is important to observe that the structure of an EDD in Def.~\ref{def:edd} mirrors that of the DNN program in Def.~\ref{def:dlp}. This is intentional: it is the goal of \oursys{} that the output of energy accounting reflects the logical structure of the DNN program. 

Most DL programs rely on a very limited set of tensor operations, i.e., the sizes of $\mathbb{TN}$ and $\mathbb{TO}$ are small. For example, \bert{} is built on top of a limited number of tensor operations, such as vector multiplication (\texttt{MATMUL}) and tangent gradient (\texttt{TANHGRAD}). % \albert{} is built on top of one tensor operation, Einstein summation with name \texttt{EINSUM}. 
However, it is important to note that \emph{where} tensors appear in the DL program makes them semantically different: a vector multiplication used for computing self-attention is different from one using Einstein summation. To effectively support this difference, we introduce \emph{qualified tensor name} $(q \in \mathbb{QN})$, in the form of $\lb [c_1, \dots, c_n ]; t\rb$ for some $n \geq 0$. Intuitively, this refers to a tensor named $t$ that immediately resides in layer $c_n$, which is in turn nested in $c_{n-1}$ and so on.

%\dnote{fix XYZ, XXX etc}

\begin{example}[Qualified Tensor Name and its ShortHand]
 The QTN $\lb \texttt{[bert, encoder, layer\_0, output, dense]}; \texttt{ MatMul} \rb$ refers to tensor \texttt{MatMul} defined in the \texttt{dense} layer, which in turn is nested in the \texttt{output} layer, which is nested in \texttt{layer\_0}, and so on. From now on, we use its more mnemonic shorthand form~\footnote{This notation is popularized by file systems, and broadly, hierarchical decomposed naming systems.} \texttt{bert/encoder/layer\_0/output/dense/MatMul}.

\end{example}

% \begin{center}
%     $\lb \texttt{[bert, encoder, layer\_0, output, dense]}; \texttt{ MatMul} \rb$
% \end{center} 

With the QTN, we can now ``flatten'' the EDD to produce a representation more friendly for tensor-level comparison of energy consumption:

\begin{definition}[Tensor Energy Footprint]
We define a \emph{tensor energy footprint} (TEF) $F$ as a function $TEF: \mathbb{QN} \rightharpoonup \mathbb{ENERGY} $.
\end{definition}

Given a program $P$, the transformation between EDD and TEF is simple, to be defined in \S~\ref{sec:algo_spec}.

\subsection{Algorithm Overview}

\paragraph{Design Challenges}

%Our discussion in \S~\ref{sec:background} presents an elegant model for NNs. Unfortunately, this all comes at a cost: a lack of accountability. Most programs use multi-threading to get high throughput. While per-thread energy accounting is understood, it does directly map to operations. 
%This leaves traditional by-thread accounting with a \emph{semantic gap}: there is no knowledge about what is done with the energy.

As we discussed in \S~\ref{sec:intro}, \oursys{} first must overcome a \emph{Semantic Gap} challenge: while it is generally known how to perform (gray-box) energy accounting in a per hardware component or per thread manner, how such systems-level energy consumption can be mapped to the structure of the DL program is an open question. Secondly, there is an \emph{Application-Systems Interface} challenge.  A typical DL program is multi-threaded, so addressing concurrency is the rule not the exception; DL programs routinely run in a \textit{heterogeneous environment} -- where there may be multiple devices, such CPUs and GPUs. This complicates accounting further as device reports their consumption separately. %through unique \textit{energy domains} -- an abstraction of hardware components. This results in a complex dialogue for the \emph{application-system interface}, requiring expert knowledge.

\paragraph{A Solution Overview}

To address the \emph{Semantic Gap} challenge, \oursys{} is designed as a runtime monitor to perform \emph{trace-based alignment}: it collects two traces of information at time intervals, and align them based on timestamps. The two conceptual traces are (1) a \emph{tensor event trace} $\tau: \mathbb{TIMESTAMP} \rightharpoonup \mathbb{QTN}$, which temporarily records the tensor activities, where each tensor is identified by its QTN; (2) a \emph{power trace}: $\pi: \mathbb{TIMESTAMP} \rightharpoonup \mathbb{POWER}$. The core algorithm is conceptually simple: \oursys{} continuously tracks what tensor events have happened, where each event has happened, and how much power the device has when it happens. Now that $\tau$ contains the semantic information, its alignment with $\pi$ bridges the semantic gap between the \emph{program} runtime and the \emph{system} runtime. Let us now overview several technical challenges: 

\begin{itemize}

\item \emph{Imperfect Alignment}: It should be known that neither $\tau$ nor $\pi$ is \emph{surjective}; in other words, some timestamps may not have a tensor event or any power reading. In a nutshell, \oursys{} is a \emph{recency}-based alignment algorithm: we align a tensor event with the most recent power reading in the time line.  

\item \emph{Multiplicity}:  $\pi$ is not \emph{injective}; in other words, there might be multiple tensor events that occur at the same timestamp. In \oursys{}, all tensor events that happen at the same time are attributed with an equal share of the energy consumption. 

\item \emph{Durable Events}: tensor events happen \emph{for a duration}. This realistic view is in contrast with our conceptual formulation above, where the $\tau$ mapping appears to indicate that the occurrence of a tensor event happens \emph{at} an instantaneous timestamp. We resolve this with a conceptual algorithm (in Algorithm~\ref{alg:smaragdine_accounting}) and an optimized algorithm (see discussion in \S~\ref{sec:algo_spec}). 

\end{itemize}

To address the \emph{Application-System Interface} challenge, \oursys{} behaves as a \textit{universal accountant}, abstracting the system's consumption as energy traces. First, \oursys{} is aware of the heterogeneity of the system, where trace alignment is performed in a \emph{per device} manner. In other words, \oursys{} operate on the \emph{device tensor event trace} $\theta: \mathbb{DEVICE} \rightharpoonup \tau$, and the \emph{device power trace}: $\epsilon: \mathbb{DEVICE} \rightharpoonup \pi$. Second, \oursys{} is aware of the concurrency of the system. %, and excludes usage from foreign applications. 
With  \emph{Multiplicity}, the tensor events that are concurrently executed on different threads residing on one device each will receive an equal share of the energy consumption of that device. \oursys{} also addresses complex systems behavior such as thread migration, where the mapping between threads and devices is updated.

\subsection{Algorithm Specification}
\label{sec:algo_spec}

\paragraph{Key Data Structures}

\begin{figure}[t]
    \begin{algorithm}[H]
        \caption{\oursys{} Data Types}
        % \dnote{I thought I was convinced yesterday that the energytrace indeed should be powertrace. Otherwise, recency-based lookup would not have made sense. I have updated the overview narrative to make it PowerTrace, and followed your description with a rewrite. Perhaps you can turn EnergyTrace back to PowerTrace? Just make sure that whatever is read from the PowerTrace should be multiplied by the duration of the event before summed up for TEF computation.  }
        \label{alg:data_types}
        \begin{algorithmic}[1] 
            \State $\textbf{typedef} \ \textsc{Ts} \ \textsc{Int}$  // time stamp 
            
            \State $\textbf{typedef} \ \textsc{Dur} \ \textsc{Int}$  // time duration

            % \State $\textbf{typedef} \ \textsc{Thread} \ \textsc{Int}$  // ID of executing thread

            \State $\textbf{typedef} \ \textsc{Device} \ \textbf{enum} \ \{$ \\ \label{alg:devices}
                \hspace{\algorithmicindent} $\textsc{CPU\_0}, \textsc{CPU\_1}, ... $ \\ 
                \hspace{\algorithmicindent} $\textsc{GPU\_0}, \textsc{GPU\_1}, ... $ \\ 
            $\}$ \label{alg:device-type}

            \State $\textbf{typedef} \ \textsc{Op} \ \textsc{Int}$ // ID of a tensor operation

            \State $\textbf{typedef} \ \textsc{Event} \ \textbf{struct} \ \{$ \\ \label{alg:event_trace_start}
                \hspace{\algorithmicindent} $\var{ts} : \textsc{Ts} $ // event start time \\ 
                \hspace{\algorithmicindent} $\var{dur} : \textsc{Dur} $ // elapsed time for event \\
                % \hspace{\algorithmicindent} $\var{thread} : \textsc{Thread} $ // executing thread \\
                \hspace{\algorithmicindent} $\var{device} : \textsc{Device} $ // executing device \\
                \hspace{\algorithmicindent} $\var{op} : \textsc{Op} $ // event operation \\
            $\}$ \label{alg:event_trace_stop}

            \State $\textbf{typedef} \ \textsc{EventTrace} \ \textsc{List}\lb\textsc{Event}\rb$ \label{alg:event_trace}
            
           \vskip 1ex

            \State $\textbf{typedef} \ \textsc{Energy} \ \textsc{Float}$  // energy in joules
            \State $\textbf{typedef} \ \textsc{Power} \ \textsc{Float}$  // power in watts

            \State $\textbf{typedef} \ \textsc{PowerTrace} \ \textsc{Map} \lb \textsc{Ts}, \textsc{Power} \rb$

            \State $\textbf{typedef} \ \textsc{DevicePowerTrace} \ \textsc{Map}\lb \textsc{Device}, \textsc{PowerTrace} \rb $

            \vskip 1ex
            % \State $\texttt{EPOCH} : \textsc{Ts}$ 

            \Function{Start}{{}} // begin accounting
            \EndFunction \label{alg:start_sampling}

            \Function{Stop}{{}} : $\textsc{DevicePowerTrace}$ // stop accounting and obtain traces since start \label{alg:stop_sampling}
            \EndFunction

            \Function{Now}{\var{T}: \textsc{Ts}, \var{TL} : $\textsc{Set}\lb\textsc{Ts}\rb$} : $\textsc{Ts}$ // retrieves the largest time stamp from a set which is smaller than a given time stamp
                \State $\textbf{return} \ \fcall{Max}(\var{ts} \in \var{TL} \ \textbf{where} \ \var{ts} <= T)$
            \EndFunction

            % \Function{Next}{\var{T}: \textsc{Ts}, \var{TL} : $\textsc{Set}\lb\textsc{Ts}\rb$} : $\textsc{Ts}$ // retrieves the smallest time stamp from a set that is larger than a given time stamp
            %     \State $\textbf{return} \ \fcall{Min}(\var{ts} \in \var{TL} \ \textbf{where} \ \var{ts} > \var{T})$
            % \EndFunction

            % \Function{IsSame}{\var{T1}: \textsc{Ts}, \var{T2}: \textsc{Ts}, \var{TL} : $\textsc{Set}\lb\textsc{Ts}\rb$} : $\textsc{Boolean}$
            %     \State $\textbf{return} \ \fcall{GetCurrInterval}(\var{T1, TL}) \ \var{=} \ \fcall{GetCurrInterval}(\var{T2, TL})$ \\
            % \EndFunction
    \end{algorithmic}
  \end{algorithm}
%  \vskip -8ex
\end{figure}

Algorithm~\ref{alg:data_types} presents the key data structures \oursys{} works with at run time. The $\tau$ and $\pi$ traces we discussed in the previous section are represented by \textsc{EventTrace} and \textsc{PowerTrace} respectively. Due to \emph{Multiplicity}, the \textsc{EventTrace} also records the duration of the tensor event (\texttt{dur}), together with its starting time (\texttt{dur}), as shown in Lines~\ref{alg:event_trace_start}-\ref{alg:event_trace_stop}. The QTN of the tensor operation is kept in the \texttt{op} field. To address the \emph{Application-Systems Interface} challenge, we also record \emph{where} in the underlying systems such event is happening, in the \texttt{device} field. Possible values are the CPU/GPU units, as shown in the \texttt{enum} definition for \textsc{Device}. 
%First we present some data structures to preserve the semantics of the data as it is transformed (). Running a neural network on some data set for training will produce an \textit{event trace} -- a log of the system execution that implements the network. Trace events contain the execution start, execution duration, and the executing device for a uniquely-labeled tensor operation  While threads do execute more than one operation, \tensorflow{} uses a worker model to manage the execution. This guarantees that no thread executes more than one operation concurrently. A potential trace is presented in Figure~\ref{fig:transformer-timeline} for two threads sharing a device.
\oursys{} provides \textsc{Start} and \textsc{Stop} methods (Lines~\ref{alg:start_sampling}-~\ref{alg:stop_sampling}) to enable an accounting session. 
%The profiling is provided through \textit{power traces} -- a log of timestamped, instantaneous power measurements. Due to the challenges mentioned in \ref{sec:challenge_2}, \oursys{} organizes the data to appropriately align the data. Each utilized device must be queried in order to account all units in the event trace. Therefore, \oursys{} \textit{scales} with the system, sampling from as many devices as it can while keeping data well-structured. We also need to consider mismatches in temporal locality. Our example timeline's event and power traces are not perfectly aligned. As a result, we must pick a rule to compare uneven regions. \oursys{} uses interval beginning, represented through the 
Due to the need for \emph{Imperfect Alignment}, utility function \textsc{Now} returns the last interval timestamp that is still smaller than the given timestamp. %We also provide \textsc{Next}, which returns smallest timestamp.
% , as well as \textsc{IsSame} to determine if two arbitrary timestamps are within the same interval.

\paragraph{TEF Computation}

\begin{figure}[t]
    \begin{algorithm}[H]
        \caption{\oursys{} Accounting}
        
        \label{alg:smaragdine_accounting}
        \begin{algorithmic}[1] 
            \State $\textbf{typedef} \ \textsc{DeviceFlatTrace} \ \textsc{Map}\lb \textsc{Device}, \textsc{Map}\lb \textsc{Ts}, \textsc{Set} \lb \textsc{Op} \rb \rb$

            \Function{Flatten}{\var{T} : \textsc{EventTrace}} : \textsc{DeviceFlatTrace} \label{alg:start_flatten}
                \State $\var{ft} \gets \textbf{new} \ \textsc{DeviceFlatTrace}$
                \For{$\var{e} \ \textbf{in} \ \var{T}$}
                    \For{$\var{t} \ \textbf{in} \ \fcall{Range}(\var{e.ts}, \var{e.ts + e.dur + 1})$}
                        \State $\var{ft}[\var{e.device}][\var{t}] \stackrel{+}\gets \var{e.op}$
                    \EndFor
                \EndFor
                \State $\textbf{return} \ \var{ft}$ \\
            \EndFunction \label{alg:stop_flatten}

            \State $\textbf{typedef} \ \textsc{TEFootprint} \ \textsc{Map} \lb \textsc{Op}, \textsc{Energy} \rb$

            \State $\textbf{typedef} \ \textsc{TsFootprint} \ \textsc{Map} \lb \textsc{Ts}, \textsc{TEFootprint} \rb$

            \Function{BuildTEFootprint}{\var{OP} : $\textsc{Set} \lb \textsc{Op} \rb$, \var{P} : \textsc{Power}} : \textsc{TEFootprint} \label{alg:start_build}
                \State $\var{of} \gets \textbf{new} \ \textsc{OpFootprint}$
                \For{$\var{op} \ \textbf{in} \ \var{OP}$}
                    \State $\var{of}[\var{op}] \gets \var{P} \ / \ \var{|ops|}$
                \EndFor
                \State $\textbf{return} \ \var{of}$ \\
            \EndFunction \label{alg:stop_build}

            \Function{GenFootprint}{\var{T} : \textsc{EventTrace}, \var{DPT} : \textsc{DevicePowerTrace}} : \textsc{TsFootprint} \label{alg:start_footprint}
                \State $\var{tf} \gets \textbf{new} \ \textsc{TsFootprint}$
                \State $\var{ft} \gets \fcall{Flatten}(\var{T})$
                \For{$\var{d} \in \fcall{Dom}(\var{t})$}
                    \For{$\var{ts} \in \fcall{Dom}(\var{ft}[\var{d}]) $}
                        \State $\var{ops} \gets \var{ft}[\var{d}][\var{ts}]$
                        \State $\var{p} \gets \var{DET}[\var{d}][\fcall{Now}(\var{ts}, \fcall{Dom}(\var{DPT}[\var{d}]))]$
                        \State $\var{tf}[\var{ts}] \gets \fcall{BuildTEFootprint}(\var{ops}, \ \var{p})$
                    \EndFor
                \EndFor
                \State $\textbf{return} \ \var{f}$ \\
            \EndFunction \label{alg:stop_footprint}

            \Function{Aggregate}{\var{TF} : \textsc{TsFootprint}} : \textsc{TEFootprint} \label{alg:start_accumulate}
                \State $\var{of} \gets \textbf{new} \ \textsc{TEFootprint}$
                \For{$\var{ts} \ \textbf{in} \ \var{x} \in \fcall{Dom}(\var{TF})$}
                    \For{$\var{op} \in \fcall{Dom}(\var{TF}[\var{ts}])$}
                        \State $\var{of}[\var{op}] \stackrel{+}\gets \var{TF}[\var{ts}][\var{op}]$
                    \EndFor
                \EndFor
                \State $\textbf{return} \ \var{f}$ \\
            \EndFunction \label{alg:stop_accumulate}
        \end{algorithmic}
    \end{algorithm}
%    \vskip -6ex
\end{figure}

Algorithm~\ref{alg:smaragdine_accounting} specifies the core monitoring algorithm which ultimately produces the TEF. Due to the technical challenge of \emph{Durable Events}, we
%\oursys{} combines event and power traces to produce \textit{operation footprints} (Algorithm~\ref{alg:smaragdine_accounting}) -- which map the tensor operations to consumed energy. Due to the device sharing discussed previously, the event trace needs to be transformed into a shape that is concurrency aware. We 
\textit{flatten} the \textsc{EventTrace}, i.e., turning it into a \textsc{DeviceFlatTrace} where the event is repeated for \emph{every} interval covered by the duration
%changing it into a mapping of devices to time series of executing operations. Due to the model used by \tensorflow{}, we known that each operation executes on an individual thread, meaning we our algorithm will work at operational granularity. However, if a multi-operation per thread model was allowed, we could extend our structures a layer deeper by using thread-mapped sets of operations. 
The \textsc{flatten} function iterates through \textsc{EvenTrace} events and add them to a \textsc{DeviceFlatTrace} from the start to the end of the event (Lines~\ref{alg:start_flatten}~-~\ref{alg:stop_flatten}). 
%Note that no operation set can have a size of zero as a result of only considering intervals. 
Finally, we account operations by iterating over each device's \textsc{DeviceFlatTrace}. The tensor operations for each time interval is assigned an equal fraction of the total energy. %Since each time interval is accounted separately, we collect our footprints into a \textsc{TSFootprint}, a mapping where per-interval TEF is record (Lines~\ref{alg:start_footprint}-~\ref{alg:stop_footprint}). 
The TEF, represented in the algorithm as \textsc{TEFootprint}, is \textsc{Aggregated} together by combining all \textsc{TSFootprint}'s.

The algorithm implemented by \oursys{} is an optimized version of Algorithm~\ref{alg:smaragdine_accounting}. In practice, if a tensor operation has a long duration, the \textsc{Flatten} process in Algorithm~\ref{alg:smaragdine_accounting} would require insert many entries in \textsc{DeviceFlattenTrace}. Our optimized algorithm keep track of the start/end timestamps of a tensor operation internally without an explicit flattening process. The specification of this optimized algorithm can be found in the repository. 

%One important consideration is that the granularity of an event trace is typically on the order of hundreds of microseconds, while that of a power trace is on the order of milliseconds. As a result, Algorithm~\ref{alg:smaragdine_accounting} may not scale well when iterating over microsecond time unit. We have designed an optimized algorithm that groups regions where no operations start or stop.

% However, most operations don't execute in such a small window of time. As a result, the footprint generation may have scaling challenges with the method described in Algorithm~\ref{alg:smaragdine_accounting}. Thus, we also present an optimized version that minimizes the number of flattened events by using dynamically interval sizes.

% By marking operations as starting or ending, we can use \textsc{Now} and \textsc{Next} to iterate over all timestamps in both data sets. As a result, we are able to exclude any timestamps that were not observed by either the NN application's tracing or \oursys{}'s sampling. When considering the trace in Figure~\ref{fig:transformer-timeline}, this optimization will only require 10 accounted intervals, while the original algorithm would need one for each time interval. Given that a typical training interval runs ~1 millisecond, this could be almost 1000 additional intervals.

\paragraph{EDD Reconstruction}

% \begin{figure}
%     \subfloat[EDD]{
%         \includegraphics[width=0.33\linewidth]{images/diagrams/design/module-footprint.pdf}
%         \label{fig:module-recomposition}
%     }
%     \subfloat[STEF]{
%         \includegraphics[width=0.33\linewidth]{images/diagrams/design/topo-identical-footprint.pdf}
%         \label{fig:topological-recomposition}
%     }
%     \caption{
%         Recompositions of the TEF for Figure~\ref{fig:transformer-timeline} into an EDD and a STEF.
%     }
%     \label{fig:accounting-footprint-recomposition}
% \end{figure}

%To further improve the explainability of energy consuption in NNs, we introduce another novel and higher-level abstraction. With a designer-guided approach, we can recompose our footprint by using modularization. The original network was built with expectations we can use as our guide. A high-level framework like \tensorflow{} preserves the locations of operations within the network. We can use this wiring to recompose our footprint by accumulating the energy of operations within a single module. 

Given a TEF, it is simple to compute a corresponding EDD. Function $\emph{T2E}(F, P)$ computes the EDD for program $P$, defined as $\lb \mathcal{N}; E \rb $, where $\mathcal{N}$ is the smallest set such that $\mathcal{N}(c_1)(c_2)\dots(c_n)(t) = F(q)$ for any $q \in \emph{domain}(F)$, $q = \lb [c_1, \dots, c_n], t \rb$, where $P = \lb \emph{N}; E\rb$. 

% \dnote{I like the EDD - STEF example, but no space. }

%A recomposition of the example transformer footprint is shown in Figure~\ref{fig:module-recomposition}, where operations are encompassed within the encoders.

\paragraph{Summarized TEF} 

Realistic DNN programs have a complex topology. This implies that a typical TEF in the real world contains numerous entries. From the standpoint of program understanding, it would be desirable if we could combine the energy consumption of ``similar'' tensors together. 

For transformer-based DNN programs such as \bert{}, an opportunity exists: while such programs consist of a large number of transformers, the internals of all transformers are \emph{self-similar}, i.e., they have indistinguishable topology~\cite{HOFMANN2003155}. This provides us an opportunity to sum up the energy consumption of all tensors that reside in self-similar transformers. In \bert{}, the boundary of a transformer is identified by module name $\texttt{layer}\_i$, where $i$ is a number that ranges the number of transformers. In other words, we can potentially sum up the following entries in a TEF:

\begin{center}
    \texttt{bert/encoder/layer\_0/output/dense/MatMul} $\mapsto$ 5 \\
    \texttt{bert/encoder/layer\_1/output/dense/MatMul} $\mapsto$ 3 \\
    ...
\end{center}

\noindent into one entry: 

\begin{center}
    \texttt{bert/encoder/transformer/output/dense/MatMul} $\mapsto$ 8 \\
\end{center} 

\begin{definition}[Summarized Tensor Energy Footprint]

A \emph{summarized tensor energy footprint} (STEF), represented as $S: \mathbb{SQ} \rightharpoonup \mathbb{ENERGY}$, maps \emph{summarized qualified tensor name} (SQTN) to energy consumption values. 
An SQTN $\emph{sq} \in \mathbb{SQ}$ has the structure of $\lb [s_1, \dots, s_n]; t \rb$ where $\{s_1, \dots, s_n\} \subseteq \mathbb{CN} \cup \{\texttt{transformer}\}$. %An ATEF is a partial function $\mathbb{SQ} \rightharpoonup \mathbb{ENERGY}$.

\end{definition}

% As shown above, c
 Computing an STEF from a TEF is simple. Given a TEF $F$, function $\emph{T2A}(F)$ compute $S$ as the smallest set s.t. $S(\emph{sq}) = {\displaystyle \sum_{\emph{sq} = \diamond(q)} F(q)} $ where $\diamond(q) = q[\texttt{layer}\_1 \mapsto \texttt{transformer}, \dots, \texttt{layer}\_n \mapsto \texttt{transformer}] $. 

In addition, \oursys{} can also produce the power counterpart of STEF, which we call Summarized Tensor Power Footprint (STPF). We elide their verbose definitions in this presentation.

%\dnote{STPF; maybe we should move that "counterpart" comment earlier here, because we never really talk about TPF. }

% In addition, we would like to compare operations across modules. We adopt a notion of \textit{topological indistinguishability}. If two networks have the same neurons and wiring, then they are indistinguishable. Similarly, each of the operations that compose the network are indistinguishable. Recall that \bert{} is constructed with a stack of transformers. Each one has the same dimensions, making them indistinguishable. We can recompose our footprint by accumulating indistinguishable operations within modules. A recomposition of the example transformer footprint is shown on the right portion of Figure~\ref{fig:topological-recomposition}, where operations are grouped by their logical function. \dnote{I think most of this can be removed now with the self-similarity discussion above. The example can stay. }

\section{\oursys{} Implementation}
\label{sec:implementation}

\paragraph{Decoupled Monitoring}

We choose to implement \oursys{} as a separate \emph{process} co-running with the monitored application. In our implementation, \oursys{} is written in \emph{Rust}. The monitored \tensorflow{} applications we evaluate \oursys{} over are Python runtimes. The inter-runtime communication is implemented through \texttt{grpc} \footnote{\url{https://grpc.io/}}, a widely used \textit{language-agnostic} remote procedure call framework. At the begin and end of the training epochs in \tensorflow{}, we use \texttt{SessionRunHook} \footnote{\url{https://www.tensorflow.org/api_docs/python/tf/estimator/SessionRunHook}} to attach callbacks to asynchronously communicate with \oursys{}. The \textsc{Start} and \textsc{Stop} methods of \oursys{} are called upon receiving the \texttt{grpc} messages.
%The Python runtime communicates with \oursys{} using a \texttt{grpc}-generated client, which provides methods for \textsc{Start} and \textsc{Stop}. %, that call the associated methods as implemented by the service. 
%\dnote{hmm, who implements start/stop? It is not in the algorithm spec.}
% \dnote{Please check. I don't know who is oursys's client in the previous writing, and who is implementing start/stop. }
%Our principled design of decoupling the monitoring logic from the monitored application is motivated by a number of considerations. First, while Tensorflow applications are hitherto predominantly implemented in Python, this is a fast developing field where library implementations in JavaScript and Android Java~\cite{tensorflow} for browser and mobile systems. 
Decoupled monitoring enables a more language-neutral approach toward the monitored application, anticipating the diversity of future ML applications, which may be written in other languages. 
%Second, if the monitoring logic and the monitored application were to live as separate threads of the same runtime, the monitored application would need to be recompiled to take into account the additional monitoring logic, undesirable for separate compilation. Third, more technically, if the monitoring logic were to be executed in the same Python runtime, Python's \textit{global interpreter lock} (GIL)~\cite{gil} would be in effect. This undesirable feature would have a severe impact on our design: the monitoring logic and the monitored application logic would not be able execute in parallel despite a concurrent multi-threaded design. 

\paragraph{Power Sampling}

\oursys{} samples the power consumption of all CPU/DRAM/GPU components of the underlying system. The power trace is broken down by device, as described at line \ref{alg:devices}
Algorithm~\ref{alg:smaragdine_accounting}.
%retrieved from domain metadata so it can be combined with a \tensorflow{} event trace as described in our accounting algorithm ().

Specifically, the CPU and DRAM energy consumption is obtained through Intel's \textsc{RAPL} interface. RAPL provides Machine-Specific Registers (MSR) to report the accumulative energy consumption of Intel processors, reporting for each \emph{domain} (i.e., motherboard socket) separately. Within each domain, it further breaks down the report by \emph{components}, i.e., the CPU cores, the uncore (cache, TLB, etc), and the DRAM regulator. The MSR is read through \texttt{powercap}, a Linux power management module where MSR values are exposed as a psuedo-filesystem. 
%\oursys{} samples all available domains/components of the underlying system, and augment each sample with the sampling timestamp, the domain ID, and the component ID.

%an OS accounting unit for cycles. The application is accounted by assigning each task a fraction based on their jiffies relative to the system total.

% \dnote{may want to say something about power domains/sockets. we also want to say it covers uncore power too like cache. we also want to say it is an Intel feature, which our system is based on. also I'm not sure what powercap is relative to RAPL}. \dnote{how about DRAM? How is it divided? } \dnote{overall, if we don't plan to explain this as a separate item in the background section, we should explain how RAPL works in detail.}

% \dnote{need discussion on Eflect. THis is very unclear. Tell readers what this problem really is. Then tell us what the solution is without using the word "virtualize"}

GPUs energy consumption is obtained through the \textit{NVIDIA Management Library} (NVML)~\footnote{\url{https://developer.nvidia.com/nvidia-management-library-nvml}}, an interface for both monitoring and managing NVIDIA devices. The NVML provides high-level querying of GPU devices, including the instantaneous power. \oursys{} samples from the NVML using the \texttt{nvml-wrapper} package~\footnote{\url{https://github.com/Cldfire/nvml-wrapper}}, which provides a thin Rust wrapper around the library.
%\oursys{} samples all available GPU devices, and augments each sample with the sampling timestamp and GPU ID. 

%The sampling implementation used to produce the provided data is written in \textit{Rust}. W
Our power sampling period is set at 4 milliseconds, which is the smallest period that we observe where power data are updated in hardware. 
%In other words, if we were to further shorten the sampling period, successive samples would produce the same reading. A shorter sampling period would also incur higher overhead, with no gain in data precision.

%can sample from the system. Sampling any quicker may result in reading before components have been updated by the system. 

\paragraph{Excluding Energy Consumption by OS and \oursys{}}

One practical concern is that the OS maintains a basic level of energy consumption, such as through daemons. In a similar vein, \oursys{} as a co-running process also incurs a small share of energy consumption itself. %In order to account for the energy consumption of the \emph{monitored} application on the CPU system, we 
We need to exclude the energy consumption resulting from the processes outside of our application, including the co-running \oursys{} process. We resort to a prior tool \textsc{Eflect} to accomplish this goal. \textsc{Eflect} was able to \textit{virtualize} the energy consumption, i.e., separating the \emph{fraction} of energy consumption due to a specific application from the rest of the system. 
%The virtualization mechanism is also sampling-based, aligned with \oursys{} design. 
In other words, for each power sample obtained by \oursys{} at line~\ref{alg:start_sampling} in Algorithm~\ref{alg:data_types}, the sample only consists of the fraction of energy consumption due to the monitored application itself. 
%Intuitively, \textsc{Eflect} is able to track the relative activeness of all processes/threads during each sampling interval --- through a low-level OS mechanism called jiffies~\footnote{\url{https://man7.org/linux/man-pages/man5/proc.5.html}} --- and accordingly divide the physical energy consumption to individual processes/threads. In addition, 
%\textsc{Eflect} is also applied to the \oursys{} runtime so it can be disentangled from our application. 
%From the standpoint of \oursys{}, \textsc{Eflect} can be viewed as a black-box service requested by --- and running within the same runtime as --- \oursys{}.
We do not virtualize the energy consumption of the GPUs. Unlike the CPUs, GPUs only execute the kernel required by our monitored application, without the background OS daemon processes. The \oursys{} process itself does not execute on the GPU.

%various steps in the training. \oursys{} client's start and stop methods are called at the start and end of training epochs.

\paragraph{Event Trace Collection}

We obtain the \tensorflow{} event trace through \tensorflow{}'s built-in profiler. The profiler monitors the application with callbacks to the start and end of all operations executed. We start the profiler through the \texttt{ProfilerHook} \footnote{\url{https://www.tensorflow.org/api_docs/python/tf/estimator/ProfilerHook}} class, an extension of the \texttt{SessionRunHook}. At the end of execution, the profiler produces an event trace using  \tensorflow{}'s \texttt{timeline} API, where events are identified by the tensor's QTN. Combined with traces from the \oursys{} hook described above, we account the epoch with Algorithm~\ref{alg:smaragdine_accounting}. % or~\ref{alg:smaragdine_optimized}.

\section{Evaluation}
\label{sec:evaluation}

We present an experimental evaluation of \oursys{} with the aim to answer the following questions:

\begin{itemize}
    \item \textbf{RQ1:} what insights can \oursys{} provide to the designers of DL applications on their energy consumption?
    \item \textbf{RQ2:} what are the precision, overhead, and scalability characteristics of \oursys{}-based energy accounting? 
    \item \textbf{RQ3:} can \oursys{} help build the downstream toolchain for understanding DL applications? 
    %be used to study a variety of DNNs, the wisdom provided by \textbf{RQ1} also be applied to topologically \textit{similar} networks?
\end{itemize}

We answer \textbf{RQ1} and \textbf{RQ2} in this section, and \textbf{RQ3} in \S~\ref{sec:variants}.
%\dnote{perhaps we will merge the first two sections into one section. As of now, both look too thin}

\subsection{Experimental Setup}
\label{sec:experiment_setup}

All experiments are conducted on a server consisting of an Intel Xeon Silver 4300 v3 2.30 GHz CPU with 20 cores, a PNY NVIDIA Quadro P5000 GPU with 2560 CUDA cores, and 64GB DDR4 of RAM. The CPU is configured with hyperthreading enabled. The machine runs with a Debian 11 OS with the default \texttt{ondemand} governor where the P-state is on. All experiments were run with \tensorflow{} 2.8 on Python 3.8. We use \bert{} through its experiment repository \footnote{\url{https://github.com/google-research/bert}, rev. {\tiny eedf5716ce1268e56f0a50264a88cafad334ac61}}. Both pre-training and fine-tuning were performed for 500 epochs with the \bert{} repository's recommended parameters:

\begin{itemize}
    \item A max sequence length of 128 words
    \item A training batch size of 32 records
    \item A learning rate of $2\times e^{-5}$
\end{itemize}

%Approximately 500 epoch's worth of timelines are collected for each experiment. \dnote{500 or 1000? consistency? also, why approximately? } 
\bert{} is pre-trained with BookCorpus\footnote{\url{https://paperswithcode.com/dataset/bookcorpus}}, a collection of free, unpublished novels, and English Wikipedia\footnote{\url{https://paperswithcode.com/dataset/wiki-en}}, which contains annotated entries from a variety of domains. \bert{} is fine-tuned with the Corpus of Linguistic Acceptability (CoLA)~\cite{warstadt2018neural} data set. % which contains expertly annotated content from a collection of linguistics publications. 
Each experiment described in this paper is repeated 5 times. %, through which the average and standard deviation are produced.

\subsection{A Bird's Eye View of \bert{}'s Energy Consumption}

\begin{figure}[h]
    \centering
    \includegraphics[width=0.4\linewidth]{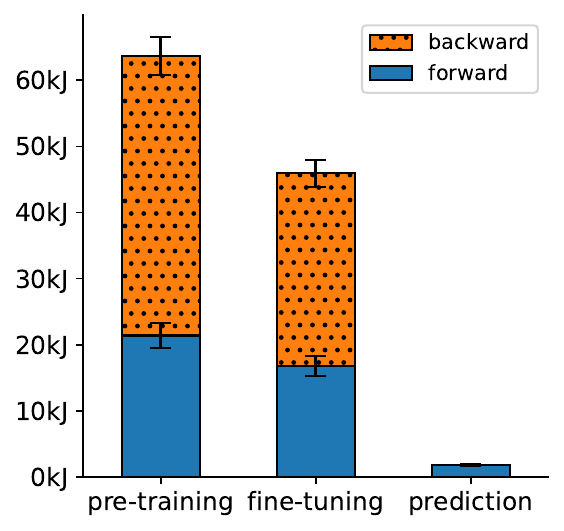}
    \caption{
        \bert{} Total Energy Consumption (The forward and backward passes are shown as stacked bars. Throughout the paper, the whiskers show the standard deviation.)
    }
    \label{fig:bert-stage_comparison}
\end{figure}

Recall from \S~\ref{sec:background} that \bert{} operates in three stages --- pre-training, fine-tuning, and prediction. \oursys{} is capable of performing energy accounting for all 3 stages. For pre-training and fine-tuning, we further separate each into the forward and backward passes of the execution, following our discussion in \S~\ref{sec:background}.

%a neural network can be run in inference mode and learning mode. Furthermore, p
%Prediction is only inference while training is inference plus learning. Therefore, breaking the accounting into these stages can facilitate our understanding of \bert{}.
%Using the labels described in Sec~\ref{sec:implementation}

Fig.~\ref{fig:bert-stage_comparison} provides a high-level comparison on the energy consumption of the 3 stages. Here, pre-training is shown as the most expensive, consuming over 60 KJs, vs. fine-tuning's 45KJ consumption. This observation is aligned with our understanding of \bert{}: pre-training updates more neurons than fine-tuning. Prediction consumes very little energy, around 2KJs. Prediction does not require the machinery of training, such as batching and iterative execution. As a result, we expect prediction to be the cheapest task.

Since pre-training is the largest consumer, as well as the first step in building a model, we will present our results of energy accounting of this stage for the rest of this section. The results for other stages are provided in the public repository (\url{https://github.com/project-smaragdine/smaragdine}).

\subsection{Multi-Grained Energy Accounting}

\begin{figure}[h]
    \centering
    % \vskip 20ex
    \subfloat[]{
        \includegraphics[width=0.5\linewidth]{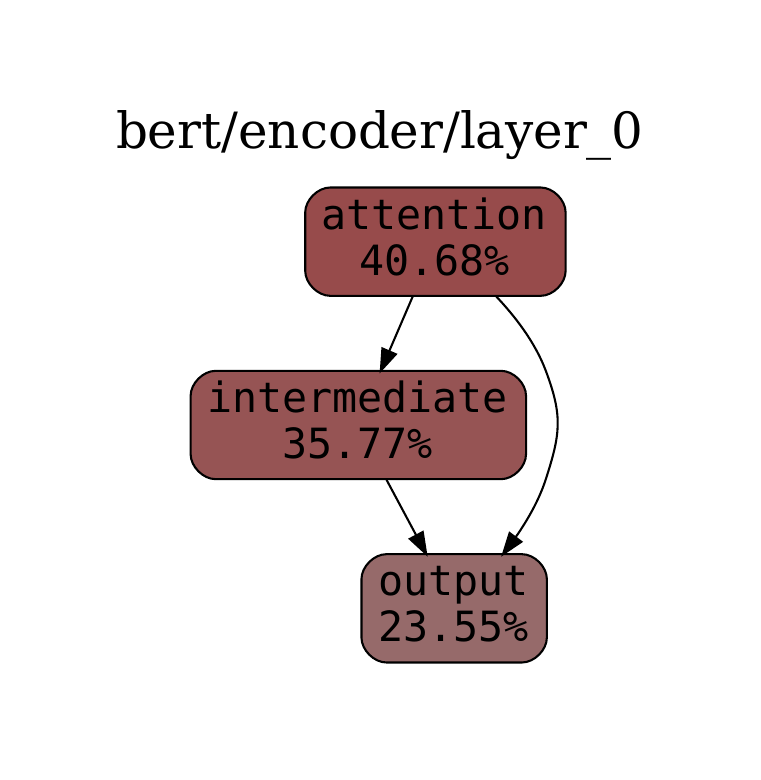}
        \label{fig:bert_encoder_layer_0-module_accounting-a}
    }
    \subfloat[]{
        \includegraphics[width=0.5\linewidth]{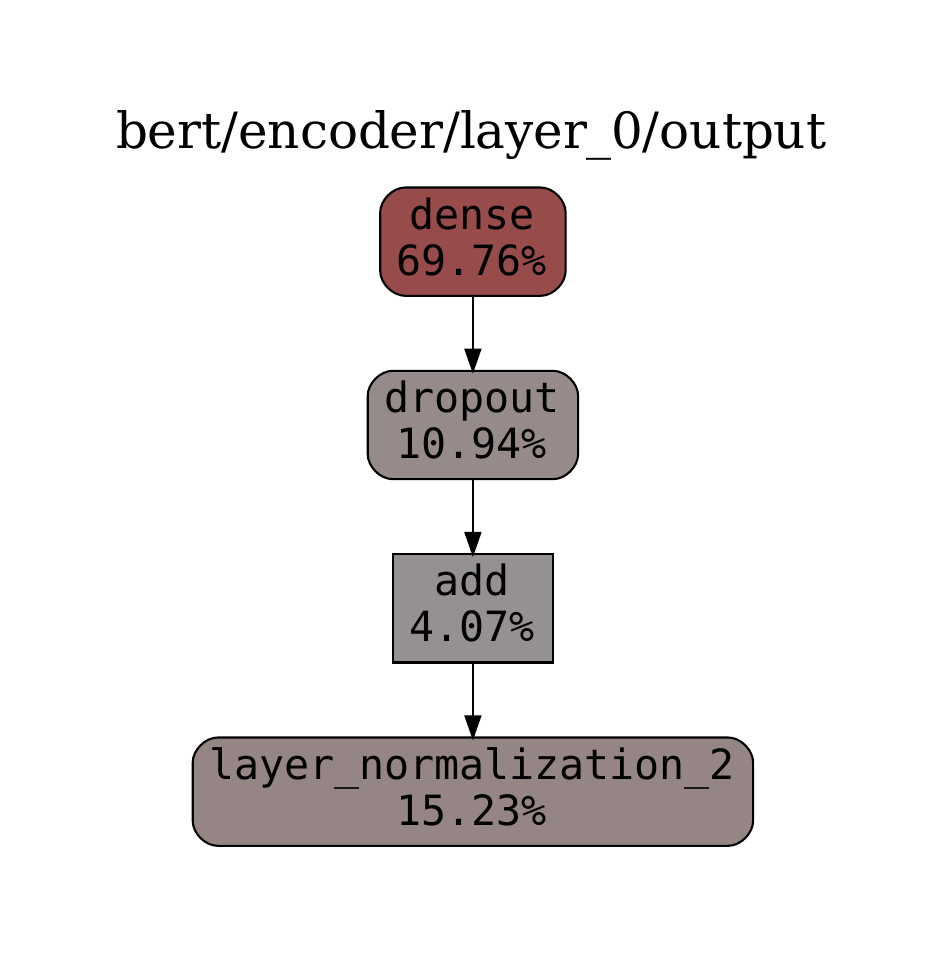}
        \label{fig:bert_encoder_layer_0-module_accounting-d}
    }

    \subfloat[]{
        \includegraphics[width=0.5\linewidth]{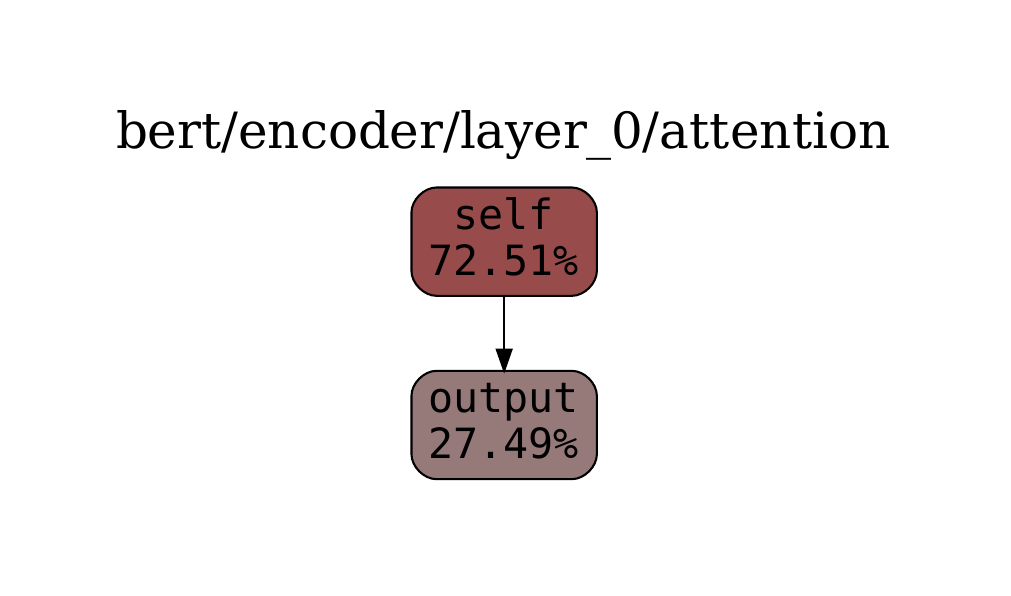}
        \label{fig:bert_encoder_layer_0-module_accounting-b}
    }
    \subfloat[]{
        \includegraphics[width=0.5\linewidth]{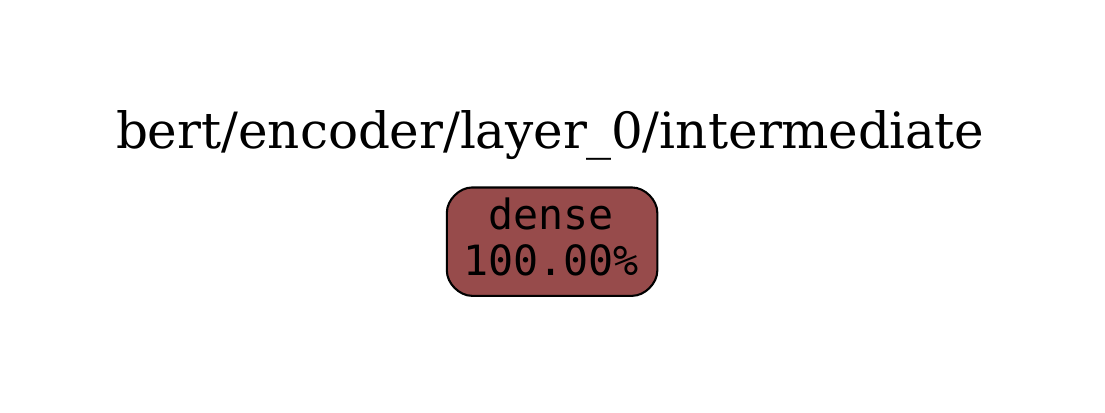}
        \label{fig:bert_encoder_layer_0-module_accounting-c}
    }
    \caption{
        Transformer-Level EDDs for the Forward Pass of \bert{}'s First Transformer (\texttt{bert/encoder/layer\_0}) and its Nested Layers.
    }
    \label{fig:bert_encoder_layer_0-module_accounting}
\end{figure}

\oursys{} is able to report energy consumption of a DL program following the logical structure of its hierarchical decomposition:

\begin{itemize}
\item \emph{(whole) program}-level accounting. At the top level, \oursys{} can provide an overview of how the energy consumption of the program is distributed among its top-level layers. For example, Fig.~\ref{fig:bert-network_accounting} is an EDD that shows the top-level view of \bert{} energy consumption. %At this scope, each unit within the diagram is a submodule of \bert{}.

\item \emph{(composite) layer}-level accounting. \oursys{} can show the energy consumption of a composite layer through nested EDDs. For example, Fig.~\ref{fig:bert-network_accounting} shows the EDD of \bert{}'s encoder layer. %At this scope, each unit within the diagram is a transformer or an intermediate tensor operation of the encoder.
%\item \emph{transformer}-level Accounting. 
Fig.~\ref{fig:bert_encoder_layer_0-module_accounting} shows a set of hierarchical views of the first transformer --- a composite layer in the encoder --- and its components. %At this scope, each unit within the diagram is another layer (either composite or tensor). % sub-transformer modules.

\item \emph{tensor}-level accounting. At this fine-grained level, \oursys{} reports the energy consumption of tensors, i.e., the ``leaves'' in the hierarchical decomposition structure. Fig.~\ref{fig:energy-operation_accounting} shows an STEF that provides a summarized view where the energy consumption of tensors is ranked. %as an energy ranking by topologically indistinguishable tensor operation. %At this scope, each unit within the ranking is a tensor operation.

\end{itemize}

\subsubsection{Program-Level and Top-Layer-Level Accounting}

\begin{figure}[h]
    \centering
    \subfloat[EDD for \texttt{bert}]{
        \includegraphics[width=0.5\linewidth]{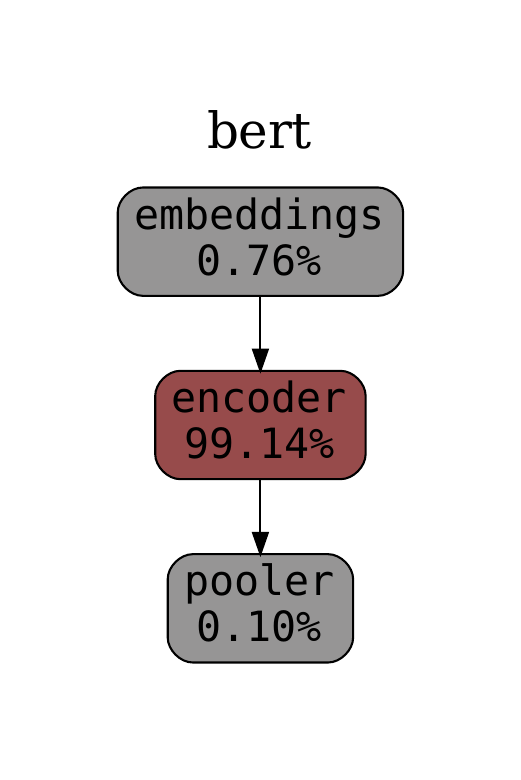}
    }
    \subfloat[EDD for \texttt{bert/encoder}]{
        \includegraphics[width=0.5\linewidth]{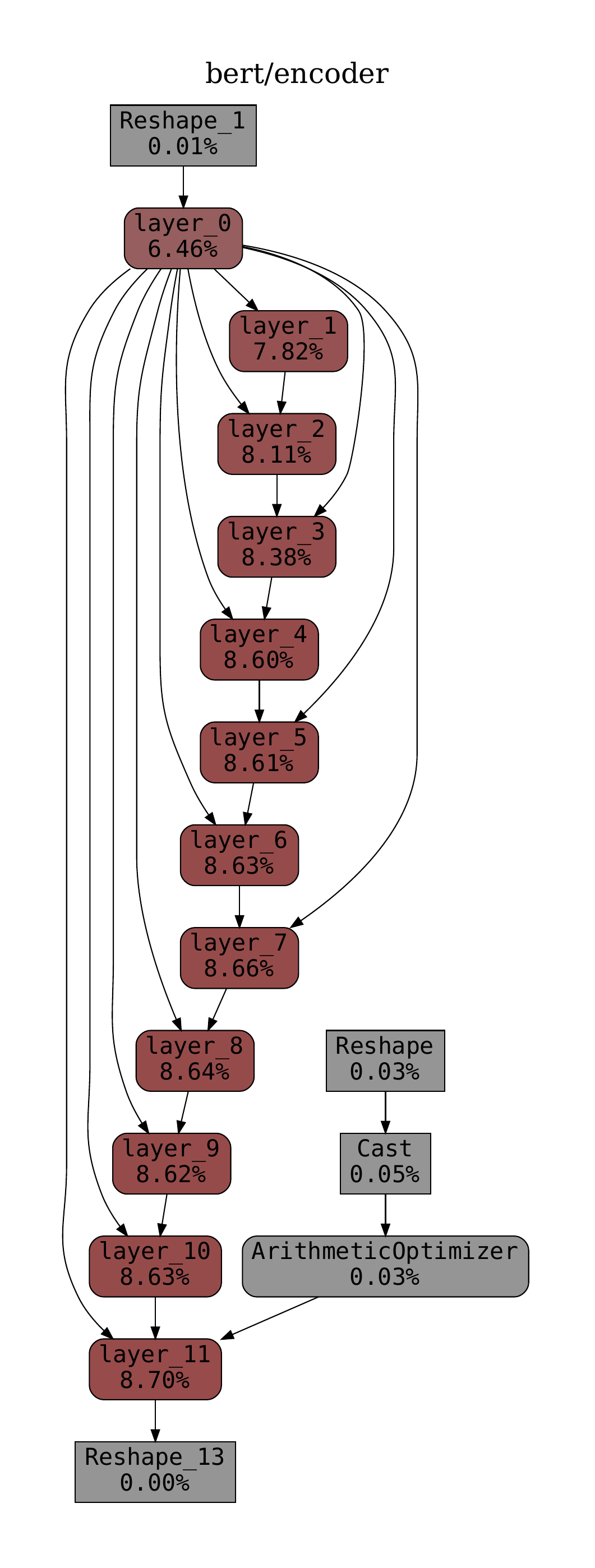}
    }
    % \subfloat[Power]{
    %     \includegraphics[width=0.5\linewidth]{images/diagrams/evaluation/module_accounting-power-forward-bert.pdf}
    % }
    \caption{
        Program-Level and Encoder-Level EDDs for \bert{}'s Forward Pass during Pre-Training (The left figure shows the top-level view, and the right figure shows the encoder layer view. The number in each box is the normalized energy consumption of the boxed unit relative to that of all boxes in the same figure. The relative values of these numbers are also represented by color intensity: a box in a more vibrant color indicates higher energy consumption. A round-edged box represents a composite layer, while a sharp-edged one represents a tensor (layer). Arrows indicate the dataflow.)
    }
    \label{fig:bert-network_accounting}
\end{figure}

In Fig.~\ref{fig:bert-network_accounting}, we first present program-level accounting for \bert{}, and the accounting for its most energy-consuming top-layer, \emph{encoder}. \oursys{} is capable of performing accounting separately for its forward pass and its backward pass. The shown EDDs show the forward pass. The results for the backward pass are deferred to our repository, with similar trends as discussed below. % It is quite similar to the inference stage, with 98\% of the energy consumed by transformers. Therefore, it will not tell us much more than the story already shown here.
%and the encoder's forward passes  as an EDD. To capture the highest-level view, we recompose the network by subsuming all operations into their submodule as described in \S~\ref{sec:implementation}. 

\begin{figure}[t]
    \centering
    \includegraphics[width=.9\linewidth]{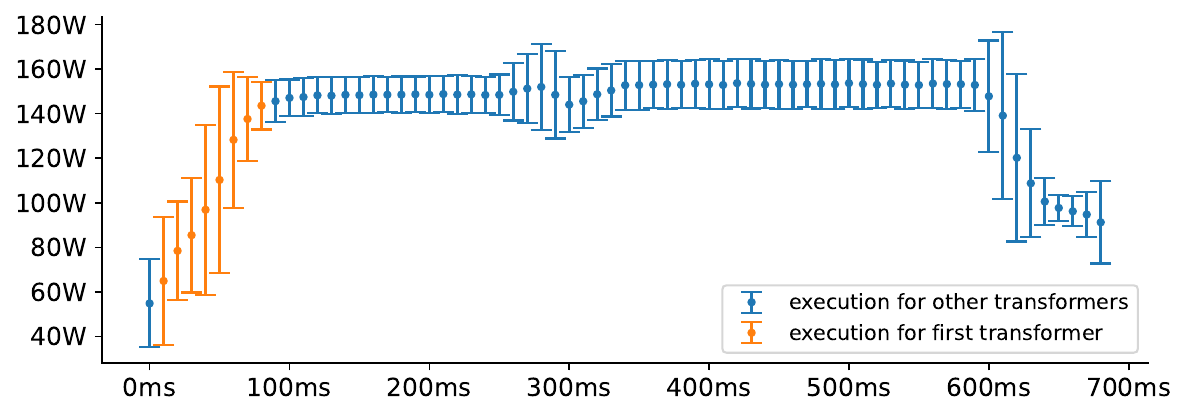}
    \label{fig:nvml-trace}
    \caption{
        \bert{} Single-Epoch Power Trace (The X-axis is elapsed time and the Y-axis is the power consumption. Each point in the figure is the average of power consumption of all training epochs at the same elapsed time. The orange data points show the time stamps when the first transformer is executed, while those in blue show the time stamps when other transformers are in execution.)
    }
    \label{fig:bert_epoch-power_trace}
\end{figure}

There are two key observations. % that can be made from these figures.
First, \emph{the encoder and its enclosing transformers dominate the energy consumption of \bert{}.} \bert{} has two main tasks: embedding and encoding. As it turns out, the latter consumes more than 99\% of \bert{} energy consumption. With the encoder, the stacked transformers again dominate the energy consumption. 
Given the central role that transformers play in language models like \bert{}, this comes with no surprise. 
%Given the complex operations that make up the transformer, this is a good sign. We expect the majority of the work to happen in the transformers, so it should also be the the largest consumer by a large margin.
Second, \emph{the transformers do not consume energy uniformly}. The transformers closer to the input are lower consumers, up to 2\% less than the average consumption of 8.25$\pm$0.61\%. The consumption rises until plateauing around 8.55\% at the fifth transformer. We speculate that this is due to the power state of the executing devices. To confirm this, we investigated the power trace of the underlying system, with results shown in Fig.~\ref{fig:bert_epoch-power_trace}. The \tensorflow{} runtime appears to schedule the transformer execution in phases, where the first transformer is executed in the earlier timestamps in each training epoch. Interestingly, the CPU/GPU system starts at a lower-power state, and only ramps up when the workload increases. As a result, operations executed in the early phases (such as the first transformer) consumes less energy overall.
%\dnote{somewhat hard to believe in the sense that for data flow programs, all are running in parallel, with new data coming in like a stream. There is no temporal difference between the 1st and the 5th in that sense. What you found is intriguing, but we just don't know why. Is there any discussion in the literature that talk about the difference between these self-similar guys? If they are identical, perhaps somehow the data underprocessing become gradually heavier? }

%\tnote{i can verify this claim with data}

% \begin{figure}[h]
%     \centering
%     \subfloat[]{
%         \includegraphics[width=0.5\linewidth]{images/diagrams/evaluation/module_accounting-energy-forward-bert_encoder_layer_0_attention_self.pdf}
%     }
%     \subfloat[]{
%         \includegraphics[width=0.5\linewidth]{images/diagrams/evaluation/module_accounting-energy-backward-bert_encoder_layer_0_attention_self.pdf}
%     }
%     \caption{
%         EDDs for Self-Attention (forward and backward passes)
%     }
%     \label{fig:bert_encoder_layer_0-module_accounting-2}
% \end{figure}

\subsubsection{Energy Accounting for Nested Layers}

Based on our earlier discussion, transformers are the largest consumers in \bert{}. We use \oursys{} to ``zoom in'' deeper in the hierarchy, to the first transformer
%~\footnote{The accounting for the other transformers are provided in our public repository.} 
(\texttt{bert/encoder/layer\_0}). Fig.~\ref{fig:bert_encoder_layer_0-module_accounting-a} shows its EDD. 

Two observations are noteworthy.
First, the \emph{attention layer} consumes a significant amount of energy; in addition, the \texttt{attention} layer's energy is primarily consumed in computing self-attention (\texttt{self}). This confirms the important role that the attention mechanism plays in \bert{}. 
Second, \emph{dense} computation dominates the energy consumption. Through the EDDs of the sub-layers in Fig.~\ref{fig:bert_encoder_layer_0-module_accounting-d}, ~\ref{fig:bert_encoder_layer_0-module_accounting-b}, and ~\ref{fig:bert_encoder_layer_0-module_accounting-c}, we can observe that the \texttt{dense} layers inside \texttt{intermediate} and \texttt{output} dominate the energy consumption. The \texttt{dense} layers are implemented as matrix multiplication, one of the most computationally intensive operations. 

% Since self-attention is a complex module, we further examine its EDD in Fig.~\ref{fig:bert_encoder_layer_0-module_accounting-2}. The self-attention layer is ``shallow'' compared to the other modules, i.e., many of its components are tensors (with sharp-edged boxes). Of these, we see that matrix-wide operations, such as matrix multiplication (\texttt{MatMul}), softmax normalization (\texttt{Softmax}), and value dropout (\texttt{dropout}), dominate, similar to our earlier observation on \texttt{dense}. This trend holds further for its nested \texttt{key}, \texttt{query} and \texttt{value} layers. For example, the \texttt{MatMul} operation dominates with a 78.52\% of energy consumption in the \texttt{key} layer. We elide these EDDs.

%consumes 78.52\% of its internal share) and \texttt{query} and \texttt{value}, which all have the same topology.

%As shown in Fig.~\ref{fig:bert_encoder_layer_0-module_accounting-2}, 
The forward pass and the backward pass exhibit similar energy behavior (EDDs in the repository), with one notable exception: 
%This hierarchical dive to the shallow self-attention motivates an exploration into the backward pass. For most EDDs, both passes are distributed similarly. However, 
the \texttt{attention} layer consumes a noticeably larger share in the backward pass, 48.19\%, than the forward pass of 40.68\%. % of the energy in the forward pass but 48.19\% in the backwards pass. 
This share difference also persists in the \texttt{self-attention} layer: the shares for the backward pass vs. the forward pass are 79.52\% vs. 72.51\%. 
%When examining both passes together, we see that \emph{consumption between the two passes are not the same}. Although some top consumers match, 
In the backward pass, gradient calculation is relatively expensive for attention layers.

\subsubsection{Tensor-Level Accounting}

\begin{figure}[t]
    \subfloat[STEF]{
        \centering
        \includegraphics[width=\linewidth]{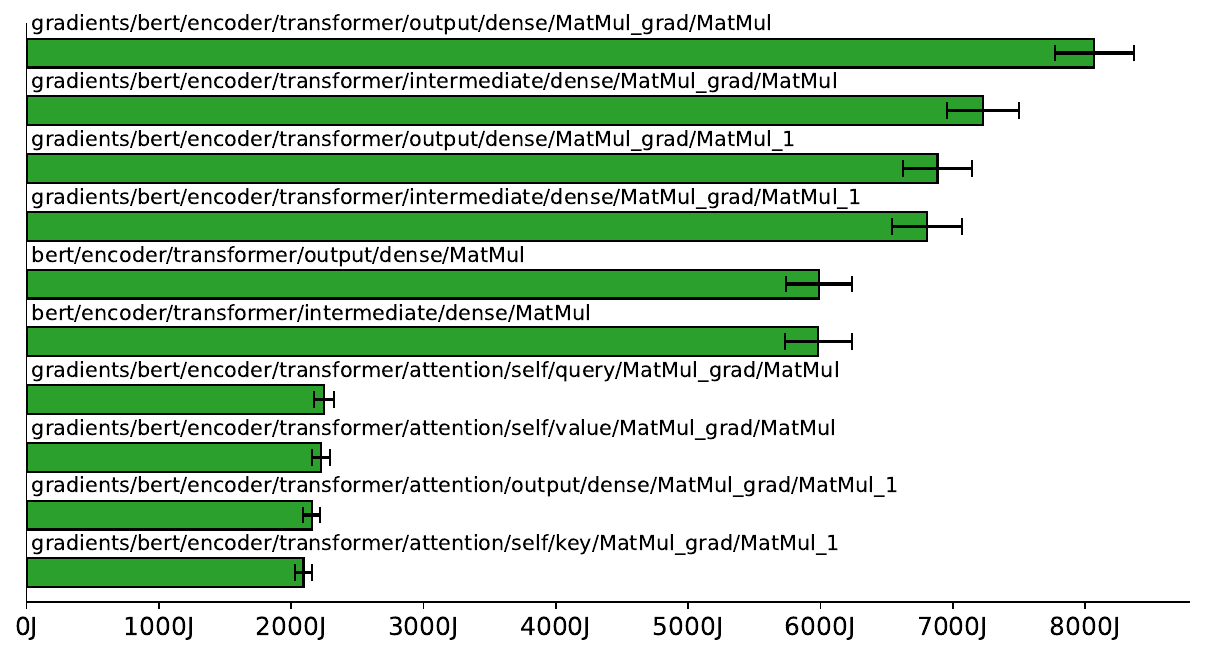}
    }

    \subfloat[STPF]{
        \centering
        \includegraphics[width=\linewidth]{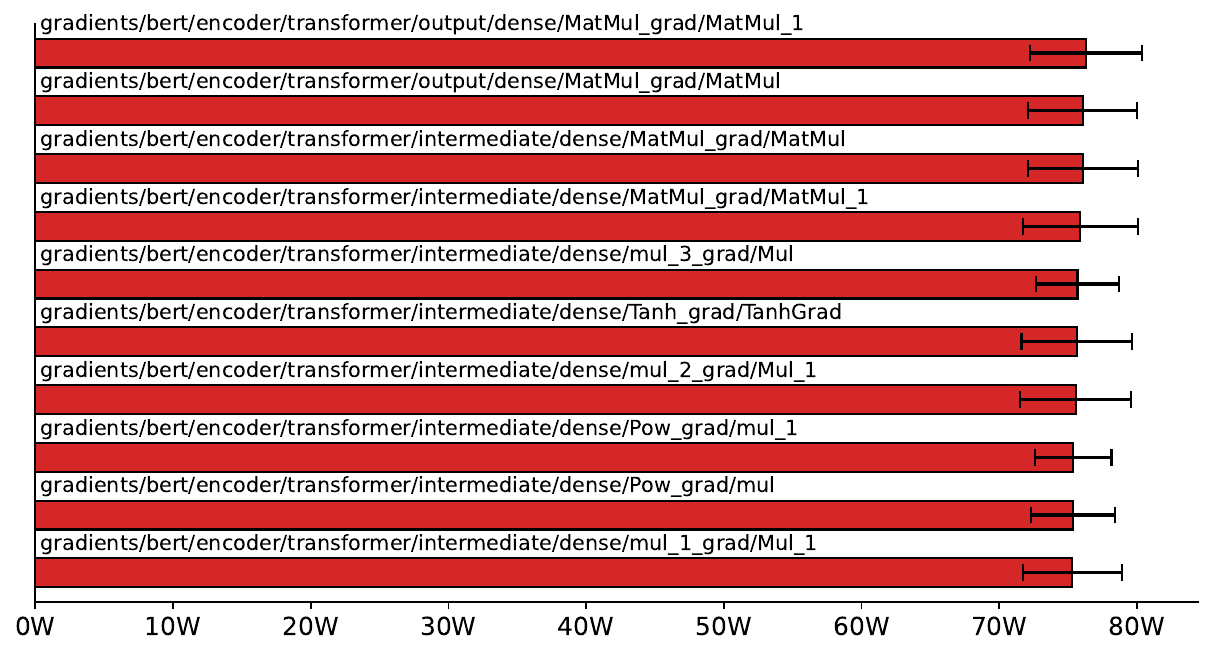}
    }
    \caption{
        Top Energy/Power-Consuming Tensors through STEF and STPF (The QTN is above each bar.)
    }
    \label{fig:energy-operation_accounting}
\end{figure}

% \begin{figure}[h]
%     \centering
%     \includegraphics[width=\linewidth]{images/diagrams/evaluation/mean_accounting_footprint.pdf}
%     \caption{
%         Instance accounting footprint of tensor operations (with normalized identities).
%     }
%     \label{fig:bert-mean-accounting}
% \end{figure}

Fig.~\ref{fig:energy-operation_accounting} presents the top-10 tensors in the form of STEF and STPF. Here, we highlight 2 observations. 
%We present top-10 energy-consuming tensor operations in Figure~\ref{fig:energy-operation_accounting}. As described in \S~\ref{sec:implementation}, we aggregate the tensor operations so the transformers with the same labels are aggregated. For instance, the integer identifier for each transformer (such as the zero in \texttt{layer\_0}) is changed to a placeholder. We present three key observations.
First, \emph{matrix multiplication} dominates the energy consumption. All top-10 energy-consuming tensors are vector multiplication across different \bert{} layers. %This is a result we have already seen in the previous accounting, so it is a good sanity check to see it again. 
%\dnote{I got rid of the first observation on matmul because earlier we wrote that's the only tensor available. Either I was wrong earlier, or this point is not needed. }
Second, \emph{the backward pass} is a much larger consumer than the forward pass. In \bert{}, all backward passes are included in the top composite layer of \texttt{gradient}. Here, 8 of 10 of the top energy-consuming tensors come from the backward pass. This is consistent with the top-level view we showed in Fig.~\ref{fig:bert-stage_comparison}, but the STEF here provides a significantly finer-grained view on which \emph{tensors} that contribute to the larger energy consumption. 
%Only two of the top ten methods are in the forward pass: the large \texttt{dense} sub-transformer modules. All the other operations are in the backward pass.
Third, the \emph{power} consumption of different tensors remains stable. Indeed, regardless of the different semantic purposes that different tensor computations serve, all share the nature of matrix multiplication (\texttt{MatMul}). As power consumption is strongly correlated with the nature of the computation itself, the power remains similar for all \texttt{MatMul} tensors. 

%\emph{the distribution of energy by operation is consistent with our previous observations}. Our shallow modules, such as \texttt{dense}, are large, single consumers, while the longer composed \texttt{attention} are multiple smaller consumers.

\subsection{Intrinsic Characteristics}
\label{sec:characteristics}

\subsubsection{Precision}

For all sampling-based systems, the precision of the results may be impacted by the sampling design itself, such as how many and how often samples are taken. 
%\dnote{I wonder if we are genuinely a sampling approach. That's for another discussion though. I guess we should at least make it clear in the design section that power is "sampled." }
We evaluate the precision of \oursys{} in two metrics:  \emph{accounting similarity with sample sparsing} (ASSS) and \emph{accounting similarity with sample widening} (ASSW). Common to the two metrics is the notion of \emph{accounting similarity}, defined as the Pearson Correlation Coefficient (PCC) of two STEFs resulted from two instances of \oursys{}, indexed by the tensor IDs. We show an example in Fig.~\ref{fig:accounting-similarity}. Intuitively, a higher value of accounting similarity implies two accounting results demonstrate more similar trends. 

% \begin{figure}
%     \includegraphics[width=\linewidth]{images/diagrams/precision/transformer-coarse-timeline.pdf}
%     \caption{
%         Execution timeline for a transformer with a slower energy sampling period than the timeline presented in Figure~\ref{fig:transformer-timeline}.
%     }
%     \label{fig:transformer-coarser-timeline}
% \end{figure}

% \begin{figure}
%     \includegraphics[width=0.75\linewidth]{images/diagrams/precision/accounting-similarity.pdf}
%     \caption{
%         Accounting similarity between the accumulated footprints produced from the timeline shown in Figures~\ref{fig:transformer-timeline}.
%     }
%     \label{fig:accounting-similarity}
% \end{figure}

\begin{figure}[t]
    \centering
    \begin{tabular}{ l | r | r }
        & \textbf{STEF$_1$} & \textbf{STEF$_2$} \\ 
        \hline
        Input Embedding & 21J & 15J \\ 
        Output Embedding & 21J & 15J \\
        Input Attention & 54J & 23J \\
        Output Attention & 54J & 23J \\
        Input Add \& Norm 1 & 118J & 59J \\
        Output Add \& Norm & 118J & 59J \\
        Input Feed Forward & 184J & 92J \\
        Input Add \& Norm 2 & 100J & 50J \\
        Output Feed Forward & 200J & 100J
    \end{tabular}
    $\rightarrow$ 0.9958
    \caption{Accounting Similarity (\textbf{STEF$_2$} has half the sampling rate of \textbf{STEF$_1$}. The PCC of the two STEFs is shown to the right.)}
    \label{fig:accounting-similarity}
\end{figure}

\begin{figure}
    % \includegraphics[width=\linewidth]{images/diagrams/precision/batch-pcc.pdf}
    % \caption{
    %     Precision metrics for ASSW through comparison of accumulated batches.
    % }
    % \includegraphics[width=\linewidth]{images/diagrams/precision/.pdf}
    % \label{fig:assw}
    \centering
    \subfloat[ASSS]{
        \includegraphics[width=0.5\linewidth]{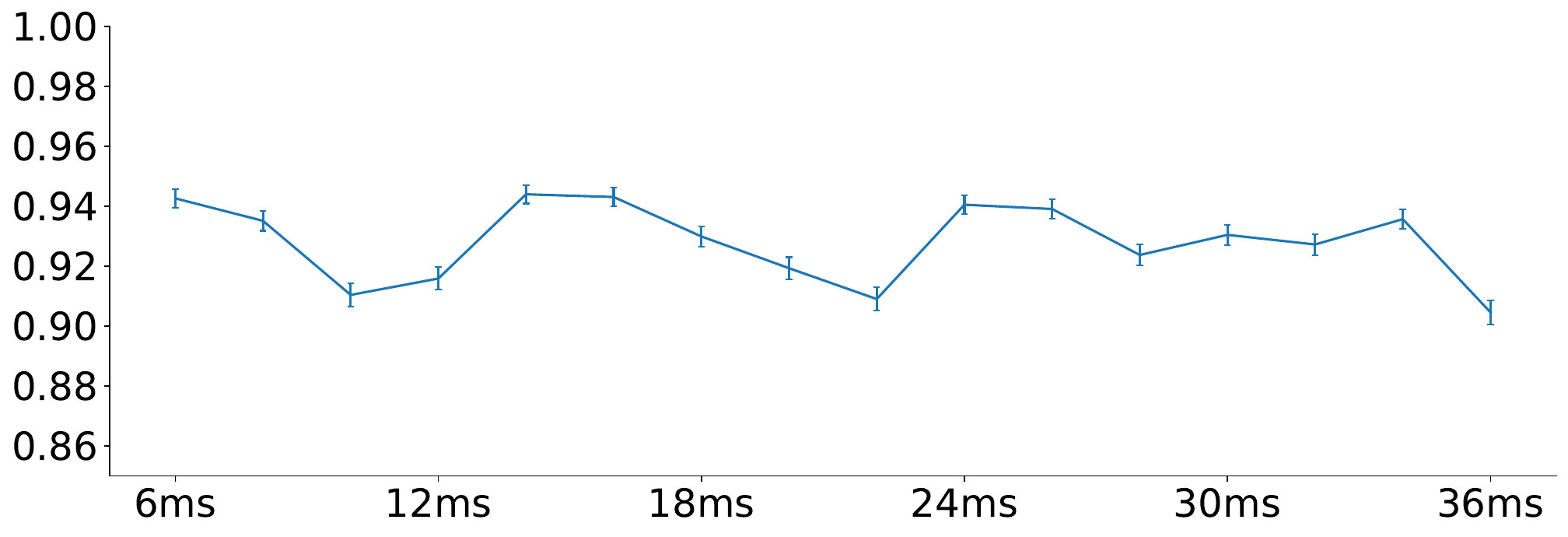}
        \label{fig:asss}
    }
    \subfloat[ASSW]{
        \includegraphics[width=0.5\linewidth]{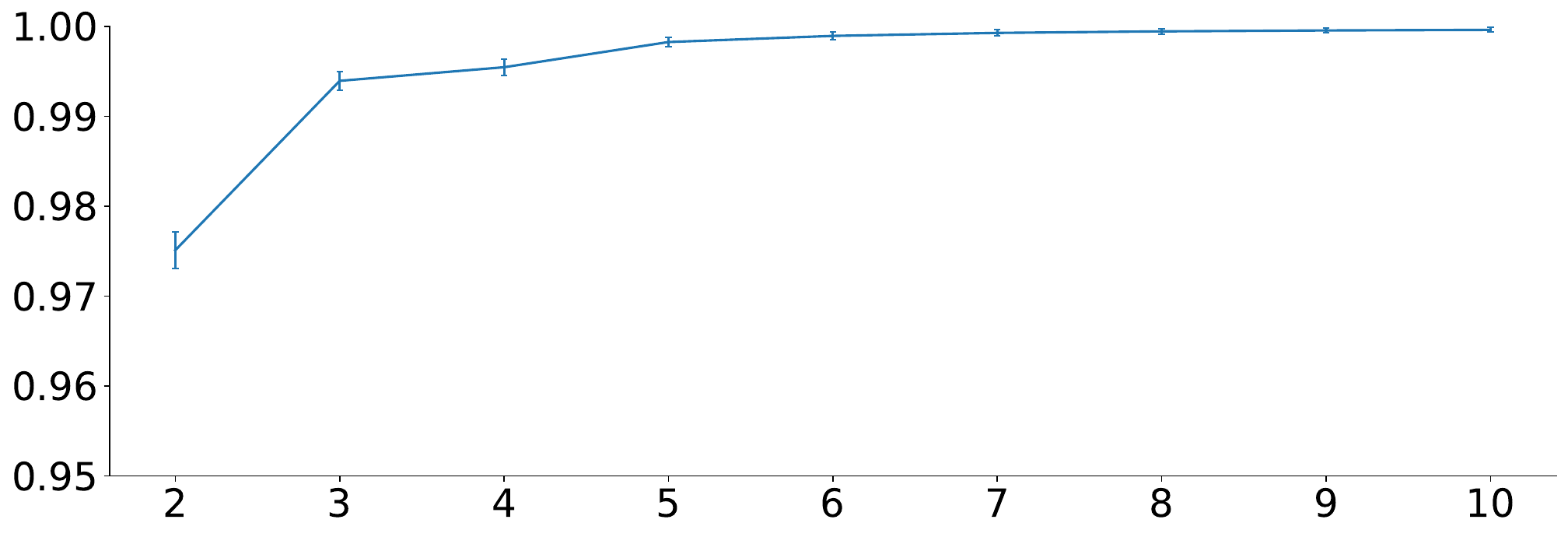}
        \label{fig:assw}
    }
    \caption{
        ASSS and ASSW (In the left figure, the X-axis is the sampling period and the Y-axis is the ASSS value. In the right subfigure, the X-axis is the number of experiments, and and the Y-axis is ASSW value. The PCC is computed between the STEF of \oursys{}'s default setting, and that of the setting in the X-axis.) 
    }
    \label{fig:precision_metrics}
\end{figure}

ASSS is built upon the intuition that the ground truth is approached when \emph{the sampling rate reaches infinity}. Recall in \S~\ref{sec:implementation}, power consumption is sampled at four milliseconds, and cannot be sampled at a higher rate due to hardware constraints. We circumvent this challenge by computing the accounting similarity when the sampling interval is further \emph{lengthened}. This counter-intuitive idea is rooted on how discrete systems approximate continuous values: if we view the result when the sampling rate is infinity as the \emph{limit}, the shape of the trajectory where the limit is approached offers clues on the error of the approximation. The results are shown in Fig.~\ref{fig:asss}. The similarity oscillates between 0.90 and 0.94. Overall, the curve forms a ``plateau'': further increasing the sampling rate would likely offer little benefit in changing the trend exhibited in the STEFs.
% at smaller periods. 

%and eventually stabilizes around 0.92 at 28ms. %This may be due to a mismatch between the aligned timelines as demonstrated in Fig.~\ref{fig:accounting-similarity}.

% , shown in Figure~\ref{fig:transformer-coarser-timeline}. We compare the

% The offset of the sample results in both differences in the measured power in an interval as well as overlap of operations within an interval. For example, while the first sampling interval in our fine-grained timeline required breaking apart the attention operations, the coarser-grained interval completely encapsulates both the embedding and attention operations. As a result, the distribution of power changes, with 15W being assigned to embeddings and 23W to attention. In order to fairly compare the two timelines, we can compare the timeline's accumulated footprints. Using a pearson's correlation, we compute the similarity between the timelines, as shown in Figure~\ref{fig:accounting-similarity}.

The intuition behind ASSW is that the ground truth of a sampling-based algorithm can be approached when \emph{the number of samples reach infinity}. Recall that each \oursys{} experiment is repeated in 5 runs (see \S~\ref{sec:experiment_setup}). We now compute ASSW by relating the STEF generated when 2, 3, 4, 5, etc, experiments are conducted, i.e., in 10, 15, 20, 25 runs. The results are shown in Fig.~\ref{fig:assw}. The similarity is high, between 0.97 and 0.99. % for two runs and quickly plateauing at 0.99 after 5 batches.

% \begin{figure}
%     \includegraphics[width=\linewidth]{images/diagrams/precision/period-pcc.pdf}
%     \caption{
%         Precision metrics for ASSS through comparison of a range of sampling periods with four milliseconds runs.
%     }
%     % \centering
%     % \subfloat[Pearson Correlation]{
%     %     \includegraphics[width=0.5\linewidth]{images/diagrams/precision/pcc_period_precision.pdf}
%     % }
%     % \subfloat[Absolute Error]{
%     %     \includegraphics[width=0.5\linewidth]{images/diagrams/precision/abs_period_precision.pdf}
%     % }
    
%     % % \includegraphics[width=\linewidth]{images/diagrams/precision/pcc_period_precision.pdf}
%     % \caption{
%     %     Precision metrics for ASSS through comparison of a range of sampling periods with four milliseconds runs.
%     % }
%     \label{fig:asss}
% \end{figure}

\subsubsection{Stability}

% \begin{figure}
%     \includegraphics[width=\linewidth]{images/diagrams/precision-new/run-pcc.pdf}
%     \caption{
%         Convergence of accounting similarity between a single TEF and many aggregated TEFs. The x-axis is the number TEFs that were aggregated.
%     }
%     \label{fig:run-convergence}
% \end{figure}

\begin{figure}[t]
    \centering
    \includegraphics[width=0.5\linewidth]{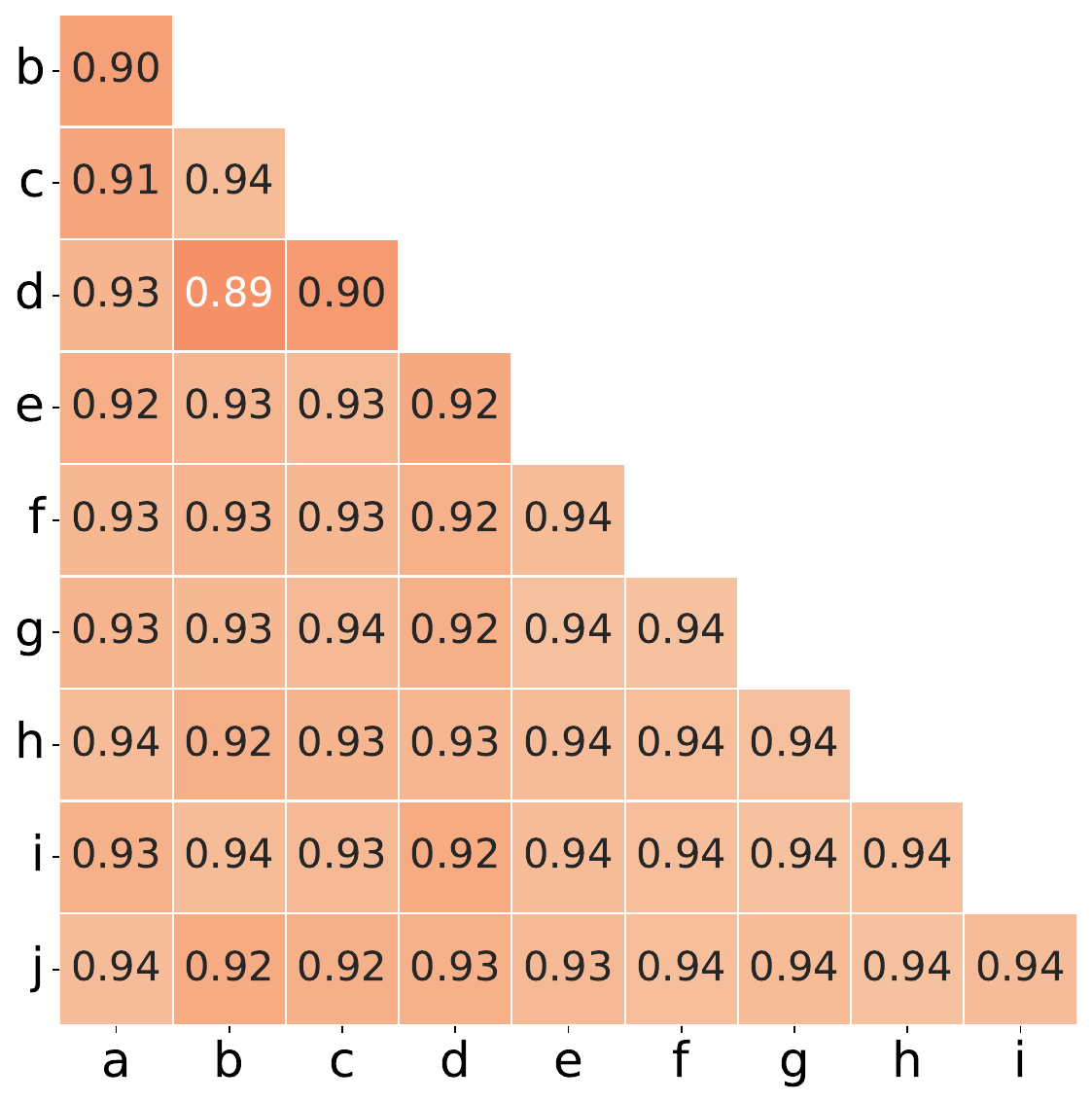}
    \caption{
        Accounting similarity across Experiments (Each box is the PCC between two experiments indexed by a-i. Each experiment is the default 5-run by \oursys{}. )
    }
    \label{fig:stability}
\end{figure}

In addition to precision, a sampling system must preserve stability: when the same experiment is repeated, a stable sampling system should produce consistent results.

To evaluate this, we can compute the accounting similarity between different experiments, shown in 
%In Fig.~\ref{fig:run-convergence}, we present the accounting similarity between different runs of the same experiment. The similarity between single runs is 0.93 and with five runs the similarity passes 0.99. In order to confirm that the produced footprints are stable, we should compare to footprints produced by a different batch. 
Fig.~\ref{fig:stability}. 
%, using the data presented in Figure~\ref{fig:assw}. A
All pairs of experiments have a high accounting similarity. Generally, a PCC greater than 0.7 is considered to be strong correlation.

%All batches have at least an average similarity of 0.91 with other batches. \dnote{I don't quite understand this. I was thinking each time we have repeated the experiments 5 times, so we should just compute the difference between those 5, with half heatmaps in the shape of ig.~\ref{fig:stability}.  Fig.~\ref{fig:run-convergence} almost feels like ASSW. }

\subsubsection{Overhead}

Finally, we quantify the overhead \oursys{} introduces to the application under energy accounting. We compare the application under \oursys{}'s accounting with the same application without accounting. %The results are shown in Fig.~\ref{fig:overhead}. 
We report a runtime overhead of $-0.052\pm0.12\%$ and an energy overhead of $0.47\pm1.10\%$. The overhead is well within the margin of errors. \oursys{} is a low overhead energy accounting system. %\tnote{this is slightly negative}

% \begin{figure}
%     \centering
%     \subfloat[Runtime Overhead]{
%         \includegraphics[width=0.75\linewidth]{images/diagrams/precision/runtime_overhead.pdf}
%     }
    
%     \subfloat[Energy Overhead]{
%         \includegraphics[width=0.75\linewidth]{images/diagrams/precision/energy_overhead.pdf}
%     }
%     \caption{
%         \oursys{} overhead produced by comparing an application under accounting with one that is not under accounting (i.e. unmonitored). The x-axis is \oursys{}'s sampling period in milliseconds and the y-axis is the absolute relative error between the accounted runtime and the unmonitored runtime. \dnote{This graph is a bit baffling because it shows values other than 4ms. I suggest we kill this figure, and only report the 4ms time/energy results in text. That will save us space too! }
%     }
%     \label{fig:overhead}
% \end{figure}
\section{Case Studies}

\label{sec:variants}

In this section, we describe two cases studies that demonstrate the usefulness of \oursys{} for supporting client studies of DL energy consumption. This section is aimed at addressing \textbf{RQ3}.

\subsection{A Comparative Study on \bert{} Variants}

As shown by the original developers of \bert{}~\cite{devlin-etal-2019-bert}, \bert{} can be configured with different hyperparameter settings. In particular, there are two important ones that impact the model topology: the number of stacked transformer layers (\texttt{L}), and the number of hidden embeddings (\texttt{H}). Our analysis in \S.~\ref{sec:evaluation} was applied to the largest model described in the original paper, i.e. $\bert{}_{\textsc{BASE}}:$ \texttt{L}=12, \texttt{H}=768.

% We can apply this accounting method to other \tensorflow{} models. Since \bert{} has spawned many variants, we choose to perform accounting on those variants. The challenge is how to do analyses across the models with different topologies.

% \begin{figure}[h]
%     \centering
%     \includegraphics[width=\linewidth]{images/diagrams/variants/transformer-comparison.pdf}
%     \caption{
%         Module accounting similarity of first transformer with\textbf{} the other transformers.
%     }
%     \label{fig:layer-comparison}
% \end{figure}

% We augment our identical topology notion by considering \textit{identical module accounting similarity}. Given two topologically identical sub-networks, the execution trace should be identical. Obviously reality is not so convenient, so we instead compute the accounting similarity between the sub-networks. We compute the pearson's correlation between the footprint of the first transformer with all other transformers (Figure~\ref{fig:layer-comparison}). The correlation is 0.92$\pm$0.02 overall, indicating that the footprints are very similar.

% \subsection{Sized-Variants of \bert{}}

% a: power pcc, b: energy pcc, c: power med, d: energy med

\begin{figure}[t]
    \subfloat[Power PCC]{
        \centering
        \includegraphics[width=0.5\linewidth]{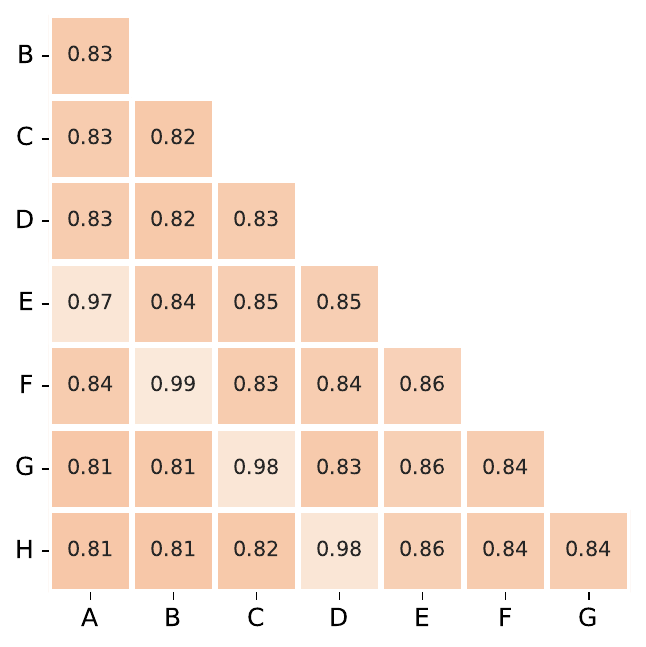}
        \label{fig:variant_power-pcc}
    }
    \subfloat[Energy PCC]{
        \centering
        \includegraphics[width=0.5\linewidth]{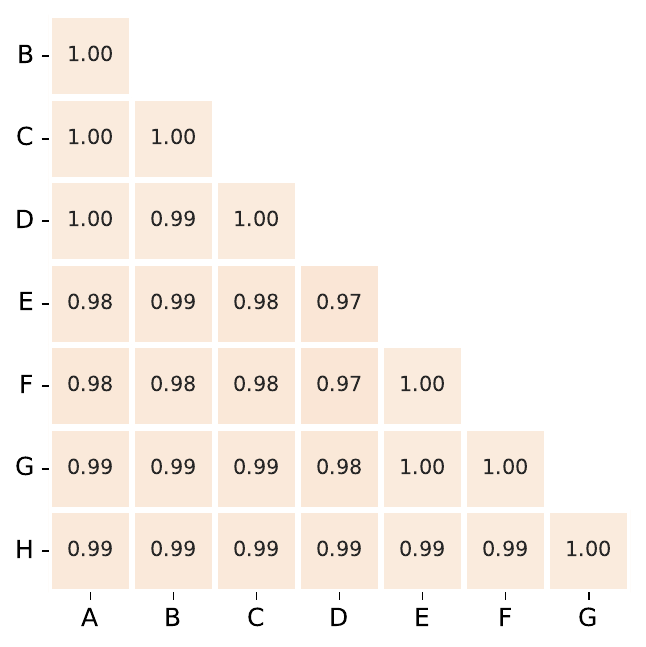}
        \label{fig:variant_energy-pcc}
    }

    \subfloat[Power MED]{
        \centering
        \includegraphics[width=0.5\linewidth]{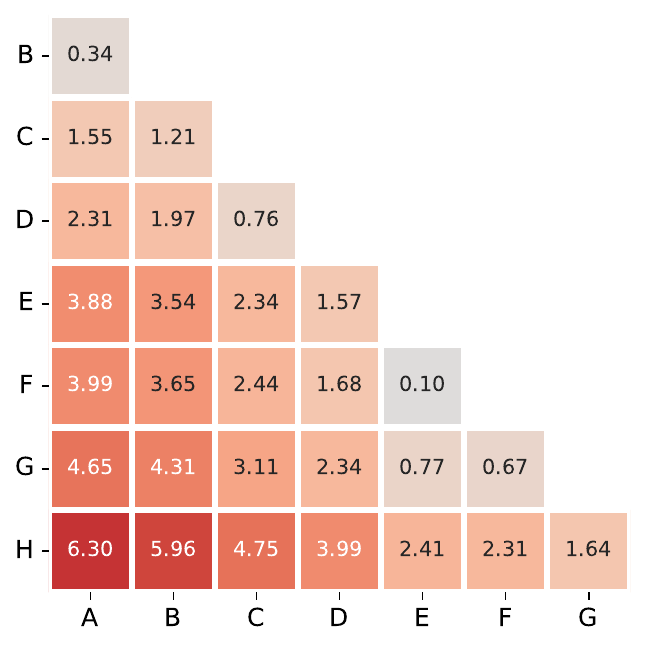}
        \label{fig:variant_power-med}
    }
    \subfloat[Energy MED]{
        \centering
        \includegraphics[width=0.5\linewidth]{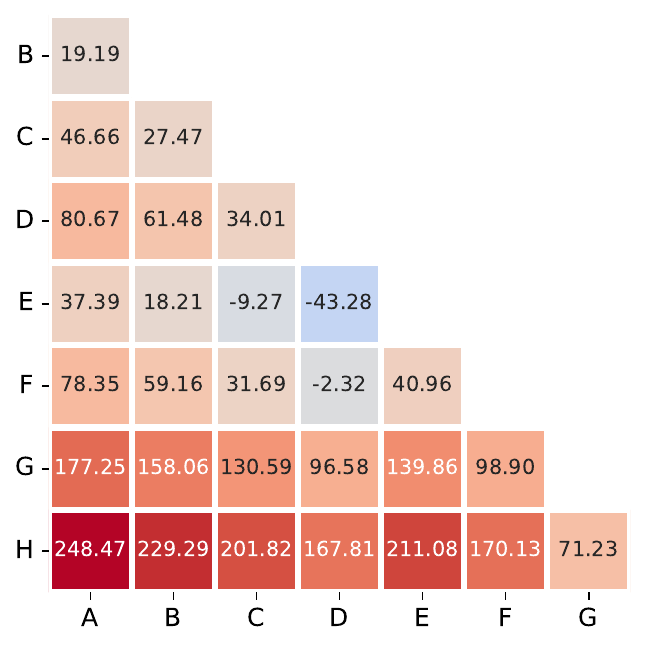}
        \label{fig:variant_energy-med}
    }
    \caption{
        A Comparative Study on \bert{} Hyperparamter Tuning (In the first/second sub-figures, each box shows the PCC between two STPFs/STEFs produced for two \bert{} variants by \oursys{}. In the third/fourth sub-figures, each box shows the mean error difference (MED) between two STPFs/STEFs produced for two \bert{} variants by \oursys{}. For MED, the variant on the Y-axis subtracts the one on the X-axis. For readability, red/blue colors indicate positive/negative numbers, and color intensity indicates the absolute values of the numbers. Each letter corresponds to a specific hyperparameter setting (\textbf{A}: [L=6, H=512], \textbf{B}: [L=8, H=512], \textbf{C}: [L=10, H=512], \textbf{D}: [L=12, H=512], \textbf{E}: [L=6, H=768], \textbf{F}: [L=8, H=768], \textbf{G}: [L=10, H=768], \textbf{H}: [L=12, H=768]).
    }
    \label{fig:variant-comparison}
\end{figure}

We use \oursys{} to generate the STEFs and STPFs for \bert{} under alternative hyperparameter settings, with the comparative results shown in  
Fig.~\ref{fig:variant-comparison}. Specifically, Fig.~\ref{fig:variant_power-pcc} and ~\ref{fig:variant_energy-pcc} show high PCC correlation across all variants in terms of both power and energy. This means that the relative standing of power/energy consumption of different tensors remains stable across different \bert{} variants. In other words, the top-consuming tensors in one \bert{} variant are likely the top-consuming tensors in the others too.

Fig.~\ref{fig:variant_power-med} and ~\ref{fig:variant_energy-med} show the results in mean error difference (MED). According to Fig.~\ref{fig:variant_power-med}, the dominating factor of power consumption is the number of hidden embeddings (\texttt{H}). All \bert{} variants with \texttt{H}=768 have a higher power consumption than their counterparts where \texttt{H}=512. For different variants with the same number of hidden embeddings, the ones with more layers consume more power. 

Fig.~\ref{fig:variant_energy-med} shows the energy trend. Interestingly, this figure does not strictly follow the trend exhibited for power consumption. It is true that when the number of layers increases, the energy consumption also increases. However, the number of hidden embeddings is no longer a deciding factor on energy consumption. For example, observe the cell between \texttt{E} and \texttt{D}, which shows the former has less (mean) energy consumption than the latter, but the former has more hidden embeddings than the latter. This is a conscious reminder to future DL program developers who are energy-conscious: both the number of layers and the number of hidden embeddings have impact on energy consumption, where neither factor may dominate.

\subsection{From \bert{} To \albert{}}

%\albert{} addressed the issue of scalability in \bert{}, where increases to the size would make usability challenging. This was done by "flattening" the topology through decomposition of the network. 
%We repeat the experiments described in \S~\ref{sec:evaluation} with 

Finally, we also apply \oursys{} to \albert{} \cite{Lan2020ALBERT:}, a variation of \bert{} with a more efficient training method.
Our experiments were conducted over the \albert{}~\footnote{\url{https://github.com/google-research/albert}, rev. {\tiny a36e095d3066934a30c7e2a816b2eeb3480e9b87}
}. The same hyperparameters are used as those in our default setting of \bert{}.

% \dnote{best have a version "head revision" is very time depenent. If no versions are possible, say the version timestamped when? }

\begin{figure}[t]
    \subfloat[STEF]{
        \centering
        \includegraphics[width=\linewidth]{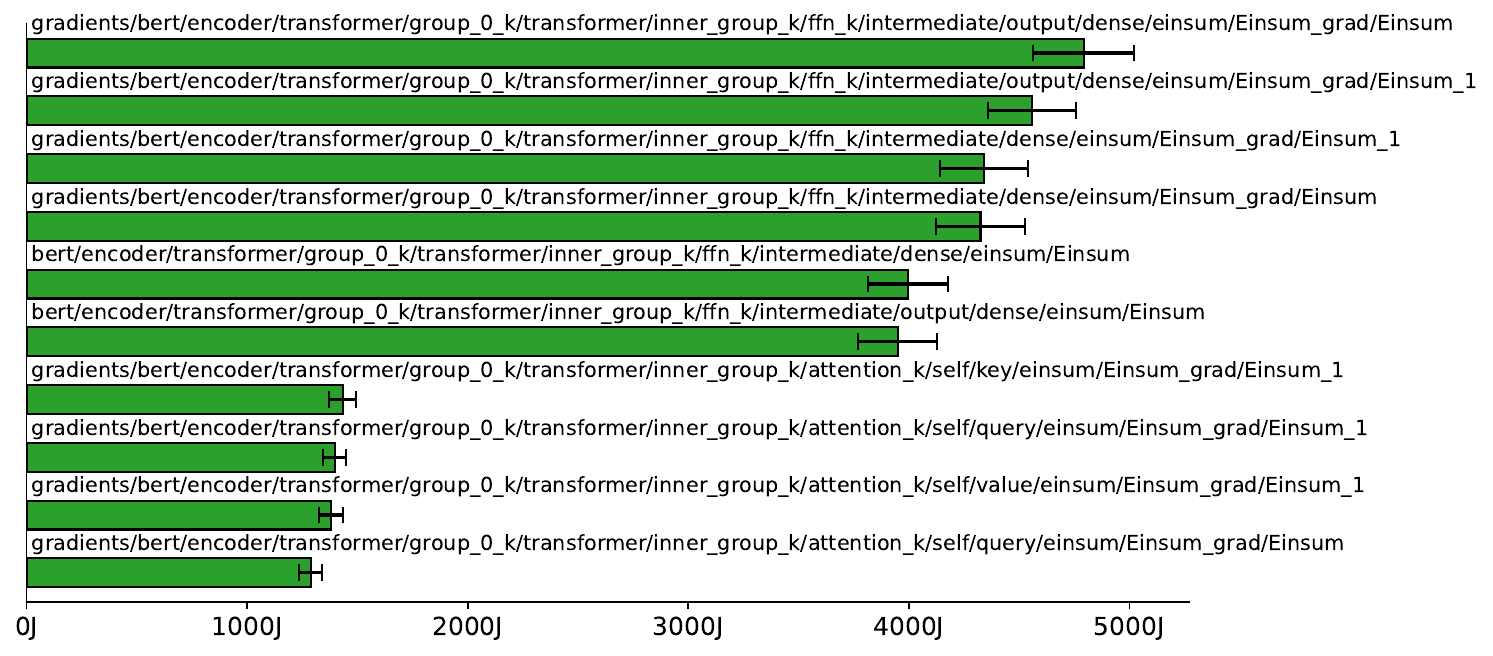}
    }

    \subfloat[STPF]{
        \centering
        \includegraphics[width=\linewidth]{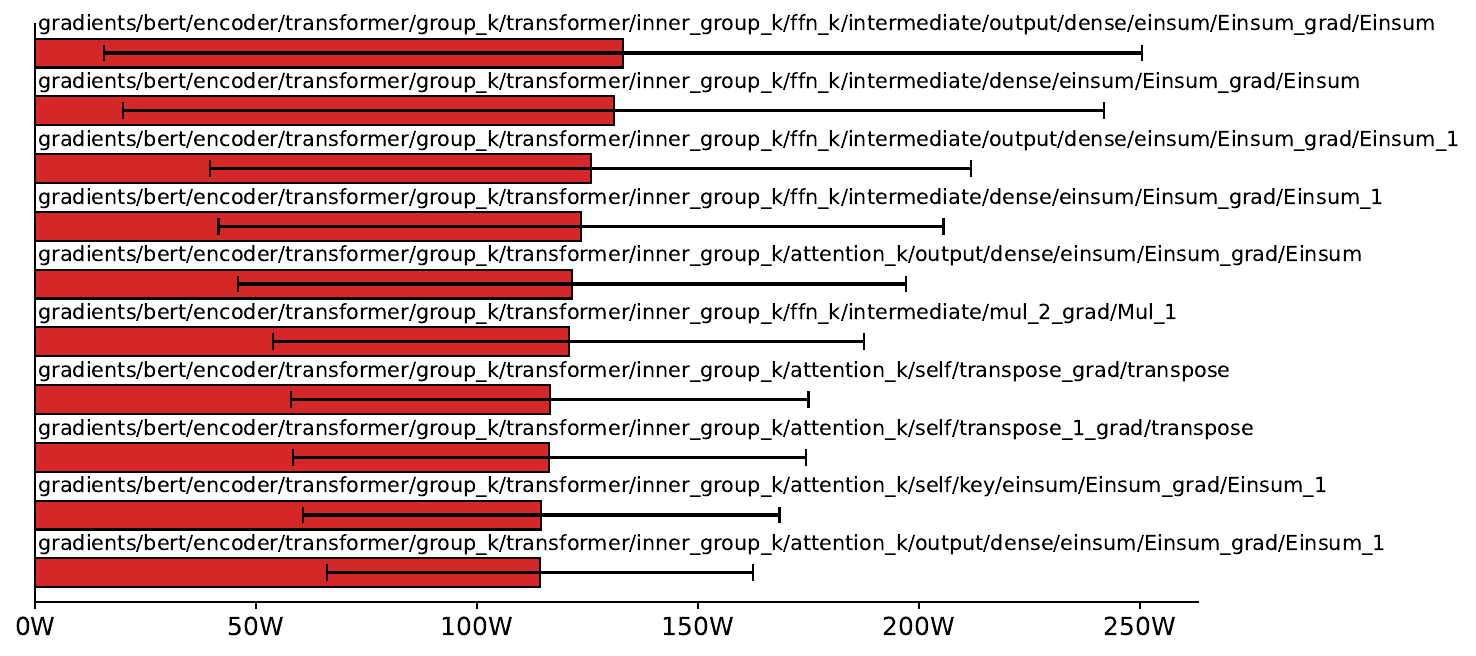}
    }
    \caption{
    Top Energy/Power-Consuming Tensors in \albert{}
    }
    \label{fig:albert-accounting}
\end{figure}

Fig.~\ref{fig:albert-accounting} presents the STEF and STPF for \albert{}. 
%, as we had done for \bert{} in \S~\ref{sec:evaluation}. 
%There are several observations we can make from these figures.
%While the name of the most energy-consuming has changed,  
%First, \emph{while the semantic naming of the topology has changed, large tensor transformations dominate energy consumption}. 
For all top-10 energy consumers, the \texttt{EINSUM} tensor is used. This refers to \textit{Einstein summation}, an index-based approach to defining tensor transformations. Indeed, matrix multiplication can also be represented with Einstein summation. From \bert{} to \albert{}, the transformation has changed from \texttt{MATMUL} to \texttt{EINSUM}, but tensor-based mathematical transformation remains dominant in energy consumption. % This indicates that tensor products dominate \albert{}'s energy consumption, much like \bert{}.

\albert{} indeed has some different consumption behavior than \bert{}. While the distribution of the energy for the STEFs is similar, the values are much smaller, almost half in some cases. In contrast, the STPF of \albert{} shows much higher power consumption --- at least 120W --- for all of the top-consuming tensors. This indicates that \albert{} is more likely to place GPUs in a higher utilization level. In addition, the standard deviations in power consumption are significantly larger than \bert{} tensors too. We speculate that it may result from changes in scheduling from \bert{} to \albert{}. Topologically, \bert{} chains the transformers in a sequential stack, i.e., the output of the $i^{\tt th}$ transformer is fed into $(i+1)^{\tt th}$ transformer as input. The topology of \albert{} however consists of multiple data pipelines. We think this design change may offer \albert{} more opportunities to execute multiple layers in parallel non-deterministically. This conjecture is also consistent with the fact that GPUs operate at a higher power in \albert{} in \bert{}. %than smaller tensor operations. These small operations individually may consume less, but together may drive the GPU into higher power states.

\section{Related Work}

%DNN energy consumption is an emerging problem hitherto addressed by a small but growing number of related work. To the best of our knowledge, \oursys{} is the first tensor-grained energy accounting system for DNNs. 

%In general, estimation of energy consumption for neural networks is very challenging, due in part to a lack of accessibility ~\cite{GARCIAMARTIN201975}. \dnote{we will rewrite this first sentence later.}
%Works like Paleo~\cite{paleo} and In addition, libraries like 

\paragraph{Energy Estimation for DL}

DeLight~\cite{delight2016} analytically models the energy cost resulted from forward and backward passes, mathematically captured through the arithmetic operations, activation functions, and propagation errors latent in the DNN. NeuralPower~\cite{neuralpower} is another analytical approach to estimating the energy consumption of a DNN given its topological details, with a mathematical model to estimate the power consumption and execution time of a DNN inference, together with parameters related to GPU scheduling such as stride size. %The work can be further traced back to a similar methodology for DNN performance prediction, such as Paleo~\cite{paleo}. 
 % and relies on profiled data to determine the parameters of the mathematical model. 
%Analytical energy estimation shares our high-level philosophy that the DNN is treated as a \emph{white box}: the topological details of a DNN can be used to understand the energy behavior of the DNN. With that said, e
Energy estimation and energy accounting are fundamentally different problems. While energy estimation offers \emph{a priori} insight of the DNN energy consumption, energy accounting is conducted \emph{a posteriori} to answer ``what happened.'' For energy estimation, a \emph{model} is assumed, e.g., how propagation and its errors are represented, how computations are kernelized, and how the GPU scheduler sets the stride size.  \oursys{} is \emph{model-less}. %it makes no assumptions on how the computation is defined, parallelized, and scheduled on a heterogeneous platform. 

%The fundeamental difference between energy estimation and energy accounting leads to different use scenarios for them: while the former is primarily used in contexts e.g., energy budgeting, the latter is useful in understanding and profiling DNNs. 

Garcia-Martin et al.~\cite{GARCIAMARTIN201975} surveyed the energy estimation approaches for ML applications, with a focus on non-DL systems. % before the DNN era. 

\paragraph{Energy Optimization of DL}

A direction that received significant interest is the energy optimization of DNNs. Domain-specific architectures and accelerators~\cite{tpus, lnpu} often deliver better energy efficiency. As a well-known example, Tensor Processing Units (TPUs) enable more performance- and energy-efficient executions for applications like \tensorflow{}. On the algorithm level, there is a long tradition in designing neutral networks with better energy efficiency. For example, SqueezeNext~\cite{squeezenext} and ChamNet~\cite{chamnet} considers energy efficiency as a key design constraint. % design goal, designing neutral networks from scratch with fewer parameters but competitive accuracy, leveraging designs such as low rank filters. ChamNet~\cite{chamnet} adapts the neutral network structure to the resource constraints placed by the execution platform. A rich and diverse number of 
Model compression techniques can often lead to increased energy efficiency, including quantization~\cite{jegou2011, gong2014, Gupta2015DeepLW}, pruning~\cite{LeCun1989OptimalBD, han2015deep}, and distillation~\cite{hinton2015distilling, huang2017like}.

Black-box systems-level approaches such as GPOEO~\cite{gpoeo} and Zeus~\cite{zeus285082} provide tools to optimize energy consumption on GPUs. For example, Zeus adaptively conducts the training of DNNs with different combinations of batch sizes and GPU power limits, and selects the more energy-optimal configurations on the Pareto curve. %Such systems approaches view the DNN as a black box. %, where the internals of the DNN is encapsulated.

Energy optimization and energy accounting go hand in hand. \oursys{} can complement existing approaches by providing a white-box view on the impact of their energy optimization, describing it in a per-layer or per-tensor manner. %If we view the evolution of a DNN as a form of manual optimization, 
\S~\ref{sec:variants} serves as examples to demonstrate how \oursys{} can help designers gain insight on energy optimization, i.e., what has really happened inside a DL program when its energy consumption is reduced/changed. 

% As a result, efforts to create more efficient networks are a critical step in achieving accountable systems without sacrificing performance. For example, the 
%SKYNET~\cite{skynet} adopts pre-training combined with post-modeling steps of problem domain information to reduce training time while retaining accuracy. 
% \dnote{removed Skynet as it is not energy optimization}

%Another option for Green AI is optimization of existing systems. With the sophisticated topologies invented in the past decade, we are seeing immense scaling of compute needs, increasing by 300,000 times from 2013 to 2029. \cite{aiandcompute}.

\paragraph{Modular DL}

%Modularization for network construction is a well established concept in machine learning. By composing the network into well-defined units, strides can be made into more powerful models that can be designed formally. As we discussed with 
\bert{} and \albert{} adopt a modular approach for model construction, whose hierarchical decomposition structure is leveraged by \oursys{} for accounting. DNN modularization is common. For example, Inception~\cite{inception1, inception2} evolves by replacing high-dimension convolutions with a sequence of small ones. Montavon et al. ~\cite{montavon2017explaining} uses Taylor decomposition of subnetworks to improve network understandability. In addition, there is emerging work in applying modularization to search for sub-networks that were not part of the design specifications ~\cite{pan:dnndecomposition, kingetsu2021neural}. %\oursys{}'s footprint can be recomposed from both kinds of definitions to further customize accounting.

\paragraph{Understanding/Profiling/Debugging DL Programs}

%As networks have rapidly scaled, designer-level insights are required in order to understand them. One of the key needs is availability of data. 
\tensorflow{} has released TensorBoard~\cite{tensorflow}, an API which collects and visualizes key metrics from a DL program runtime. Amazon similarly created the Amazon SageMaker Debugger~\cite{sagemaker} for real-time monitoring of DL applications. Both of these tools allow designers to compare training and performance metrics. PACE~\cite{pace} performs accuracy estimation over both the data set and model to identify potential issues before performing full training.
%Historically, networks were debugged manually, with designers and engineers required to identify failures. To further enable designer-friendly systems, automatic 
Debugging DL programs~\cite{mode2018, deeplocalize2021, autotrainer2021, umlaut2021, deepfd2022} is an important direction in software engineering. 
%for repairing DNNs is a big topic. MODE ~\cite{mode2018} uses a differential analysis to identify faulty features from neuron weights. AutoTrainer ~\cite{autotrainer2021} is designed to detect common training issues such as gradient failures. UMLAUT ~\cite{umlaut2021} uses a set of best practices in order to link faults to potential designer errors.

\paragraph{Non-DL Energy Accounting, Profiling, and Modeling}

For non-DL applications, energy accounting can be conducted at the levels of hardware (e.g.,~\cite{icount}), OS (e.g., ~\cite{currentcy}), and applications (e.g.,~\cite{chappie}). 
%, there is a large body of research on energy accounting in computer systems whose high-level motivations we also share. iCount~\cite{icount} accounts for energy through digital circuits. Currentcy~\cite{currentcy} and EcoSystem~\cite{ecosystem} take an accounting-by-design approach, with an operating system design where the energy consumption of sub-components are well accounted for. On the application level, Chappie ~\cite{chappie} enables method-level energy accounting for Java applications. 
%demonstrated real challenges in maintaining application calmness when profiling. Machine learning frameworks can greedily consume all resources available, making the need to minimize system impact more important.
Energy profiling~\cite{eprof,eandroid} and energy modeling~\cite{zamani, bircher,mccullough} are established directions in software engineering and computer systems. Their benefits in program understanding and debugging are well known. %Profiling approaches on mobile systems such as Eprof~\cite{eprof} and E-Android~\cite{eandroid} attribute consumed energy to different portions of the system, spanning across both software events and hardware components. Energy modeling for server systems has used performance counters~\cite{zamani, bircher} as features. However, previous modeling has been somewhat conservative, relying on a single probe as a predictor, and only sampling once per second. In addition, this modeling has trouble scaling to multi-feature and multi-threaded applications~\cite{mccullough}.

% \dnote{maybe more cites to place into these two groups? We don't need to discuss them; just cite and done.}

%Energy modeling is a very old challenge in sciences, and the impact of ML and DNNs are a very real concern. Previous work for modeling systems primarily focus on predicting instantaneous power using performance counters~\cite{zamani, bircher}. However, this modeling has trouble scaling with multi-threaded programs~\cite{mccullough}, a core behavior of a modern ML systems. This work was also conservative in sampling, with a period of once per second.

%Energy accounting is an important component of managing energy within an application. However, it is critical to not disturb the application. Work to balance performance losses while maintaining reliability through sampling is a standard practice~\cite{hirzel:2001, Whaley:2000}. Chappie ~\cite{chappie} demonstrated real challenges in maintaining application calmness when profiling. Machine learning frameworks can greedily consume all resources available, making the need to minimize system impact more important.

%\oursys{}'s awareness of the system means that it will maximize data to produce a very fine-grained trace while also retaining the concurrent context that challenges previous approaches.

\section{Threats to Validity}

First, our experiments are constructed on systems only with CPUs and GPUs; no additional accelerators are available. Hardware acceleration for DL programs is a rapidly developing field~\cite{tpus}. Our algorithm can be extended to support additional hardware, through extending the energy domains, i.e, the \textsc{Device} construct at Line~\ref{alg:device-type} in Algorithm~\ref{alg:data_types}. Second, we rely on RAPL for monitoring CPU power consumption, available primarily on Intel architectures, and on Nvidia-specific interface for monitoring GPU power consumption. This limitation can be overcome through external meters or power modeling~\cite{zamani, bircher}. Third, the \tensorflow{} runtime we currently support is in Python. While this is in sync with the majority of \tensorflow{} applications (\bert{} and \albert{} are both developed in Python), we do not have experimental evidence on the effectiveness of \tensorflow{} energy accounting in other language runtimes. In \S~\ref{sec:implementation}, we discussed decoupled monitoring, which may facilitate prototyping of our designs for other languages.

\section{Conclusion}

Tensor-aware energy accounting is a novel methodology where the accounting of energy consumption is aligned with the hierarchical decomposition structure of  nested deep neural networks defined in \tensorflow{}-based deep learning programs. Through energy distribution diagrams and tensor energy footprints, our energy accounting system, \oursys{} is capable of revealing insights on the white-box energy behavior of two widely used natural language models, \bert{} and \albert{}.

%\oursys{} is a novel energy accounting system for \tensorflow{} neural networks. \oursys{} maps event traces to system information, accounting for concurrency and locality. The operation-granular footprint produced by \oursys{} is customizable through recomposition. 

% \ourframework{} is a novel application-level software framework aiming to port energy-aware applications to a shared environment. \ourframework{} monitors the activities of the underlying system in a fine-grained manner, but exposes a minimal interface to the client application. Through energy virtualization, \ourframework{} can guide energy approximation and energy profiling frameworks to achieve their intended goals while avoiding energy entanglement. \ourframework{} is a lightweight approach that requires no modification to VM/OS/hardware, and can be adopted to commodity computer systems.

%The supplementary material contains all raw data, as well as additional figures for benchmarks not highlighted in the paper. 

%\dnote{don't forget to pack up the data} 

\bibliography{smaragdine}

\end{document}